\documentclass[10pt, a4paper]{article}

\usepackage{lrec-coling2024}
\usepackage{hyperref}
\usepackage{url}
\usepackage{times}
\usepackage{latexsym}
\usepackage{graphicx}
\usepackage{caption}
\usepackage{subcaption}
\usepackage{booktabs}
\usepackage{multirow}
\usepackage{amsmath}

\title{Analyzing the Performance of Large Language Models on Code Summarization}

\name{Rajarshi Haldar, Julia Hockenmaier}

\address{Department of Computer Science, University of Illinois Urbana-Champaign \\
         \{rhaldar2, juliahmr\}@illinois.edu\\}

\abstract{
Large language models (LLMs) such as Llama 2 perform very well on tasks that involve both natural language and source code, particularly code summarization and code generation. We show that for the task of code summarization, the performance of these models on individual examples often depends on the amount of (subword) token overlap between the code and the corresponding reference natural language descriptions in the dataset. This token overlap arises because the reference descriptions in standard datasets (corresponding to docstrings in large code bases) are often highly similar to the names of the functions they describe. We also show that this token overlap occurs largely in the function names of the code and compare the relative performance of these models after removing function names versus removing code structure. We also show that using multiple evaluation metrics like BLEU and BERTScore gives us very little additional insight since these metrics are highly correlated with each other.
 \\ \newline  }

\begin{document}

\maketitleabstract

\section{Introduction}

There is a growing interest in applying NLP techniques to tasks related to automated program understanding, generation, and retrieval, all of which promise to improve access to code. Popular tasks include  Code Summarization \cite[translating code into natural language, e.g. ][]{miceli-barone-sennrich-2017-parallel}, Code Generation, Code Completion, Code Translation and Natural Language Code Search \cite[retrieving a code snippet given a natural language query, e.g.][]{deepcs}. There is a practical need for such systems since the ability to automatically generate code snippets or doc strings or search large code bases can significantly increase the productivity of software developers. However, there is also a growing interest in developing models and datasets for these tasks within the NLP community, often driven by an assumption that code can be seen as a semantic interpretation of its natural language description.

But while current models show impressive performance, it is still important to analyze how much understanding these models have of the structure or semantics of the code. To make their code more human-readable, software developers often employ English words in the names of functions, variables, or data structures. And, although they did not evaluate the most recent web-scale large language models, the authors of MCoNaLa~\cite{wang-etal-2023-mconala} showed that the performance of code generation models drops significantly compared to English if the input is in Spanish, Japanese, or Russian.

We, therefore, ask to what extent the large language models (LLMs) that are used for tasks like code generation or summarization actually understand the semantic relation between natural language and code, and to what extent they simply rely on this superficial token similarity. 
In this paper, we attempt to address this question by analyzing the performance of large language models (LLMs) on code summarization. Our analysis aims to shed light on the following, more specific questions:  (1) to what extent do the summaries generated by these models simply consist of tokens that are directly copied from the code? (2) How much does model performance depend on the presence of function names that give away the semantics of the code? (3) To what extent do these models rely on the syntactic structure and underlying logic of the code?  

To answer the first research question, we split the examples from the dataset into different buckets depending on the token overlap between the code and the description and see how much the performance varies across those buckets. For the second and third research questions, we make several transformations to the code before feeding it to an LLM. This includes changing or obfuscating certain function names, and removing the control structures in the body of the code. We then examine their impact on code summarization performance. We observe the effect of the code in standard datasets like CodeXGLUE having informative function and identifier names with a high token overlap with their descriptions, and analyze how this affects model behavior. This token overlap between the function names and the target summary makes the task easier. Our experiments\footnote{The code for this paper will be released at \url{https://github.com/rajarshihaldar/analyze-llm-code-summarization}} show that the performance of several state-of-the-art LLMs is often due to the high string similarity of the natural language descriptions to the code they are paired with.

\section{Background}
\subsection{Large Language Models For Code}

Applying natural language processing techniques to source code has produced outstanding results. Code2vec~\cite{alon2019code2vec} showed that techniques that are used to induce semantic embeddings or vectors for natural language input also work well to represent input code snippets and predict their semantic properties. DeepCS~\cite{gu2018deep} showed that by mapping both code and natural language prompts to embeddings you could perform retrieval on code. Another breakthrough was the introduction of sequence-to-sequence (seq2seq) models for generating comments for a given input code~\cite{hu2018deep}. These initial models treat code as a sequence of tokens and were quickly followed by models that account for the structure of the code through Abstract Syntax Trees~\citep{zhang2019,wan2019,haldar-etal-2020-multi}, Graph Neural Networks~\citep{Sieper2020,ling2021,liu2021graphsearchnet}, and Graph Attention Neural Networks \cite{wang2022gypsum}.

Inspired by the success of transformer-based~\cite{NIPS2017_3f5ee243}, pre-trained large language models (LLMs) like BERT~\cite{devlin-etal-2019-bert}, XLNet \cite{NEURIPS2019_dc6a7e65}, GPT~\cite{NEURIPS2020_1457c0d6}, RoBERTa \cite{roberta}, SynCoBERT~\cite{wang2021syncobert}, and T5~\cite{raffel2020exploring} on core NLP tasks, and facilitated by the availability of large datasets that pair natural language with code, e.g. CodeSearchNet~\cite{husain2019codesearchnet}, LLMs are now commonly used for tasks that involve source code and natural language. CodeBERT~\cite{feng-etal-2020-codebert}, and its successor GraphCodeBERT~\cite{graphcodebert}, were two of the initial wave of LLMs trained on paired natural (NL) and programming language (PL) sequences. Soon other LLMs like PLBART~\cite{ahmad-etal-2021-unified} and CodeT5~\cite{wang-etal-2021-codet5} made significant gains in code summarization. At the same time, models like CodeGen~\cite{Nijkamp2022CG} and Codex~\cite{chen2021evaluating} made gains in code generation.

The current state of the art includes even larger instruction-tuned models that can perform a wide variety of tasks including ones related to code-NL, e.g.  GPT-4~\cite{openai2023gpt4}, Llama 2~\cite{touvron2023llama} and PaLM 2~\cite{anil2023palm}. Instead of being fine-tuned for a specific task, these models are trained to respond to a prompt describing a task followed by an input. This leads to improved performance on a variety of code-NL tasks. While GPT-4 has the best performance\footnote{Although it is unclear what specific data GPT-4 has been trained on, their training data is likely to include the code snippets contained in the datasets used in this paper and data contamination is known to be a significant factor in the performance of these models on coding tasks~\cite{narayanan_kapoor_2023}. 
Moreover, GPT-4 is frequently updated behind the scenes based on the data customers provide. This makes any analyses on this data non-reproducible. Due to these concerns, we have not included GPT-4 in our studies.}, some researchers have found ways to improve its output further. For example, leveraging verbal reinforcement on an LLM improves its code generation capabilities~\cite{shinn2023reflexion}.

\begin{figure}[!t]
\centering
  \includegraphics[width=0.9\linewidth]{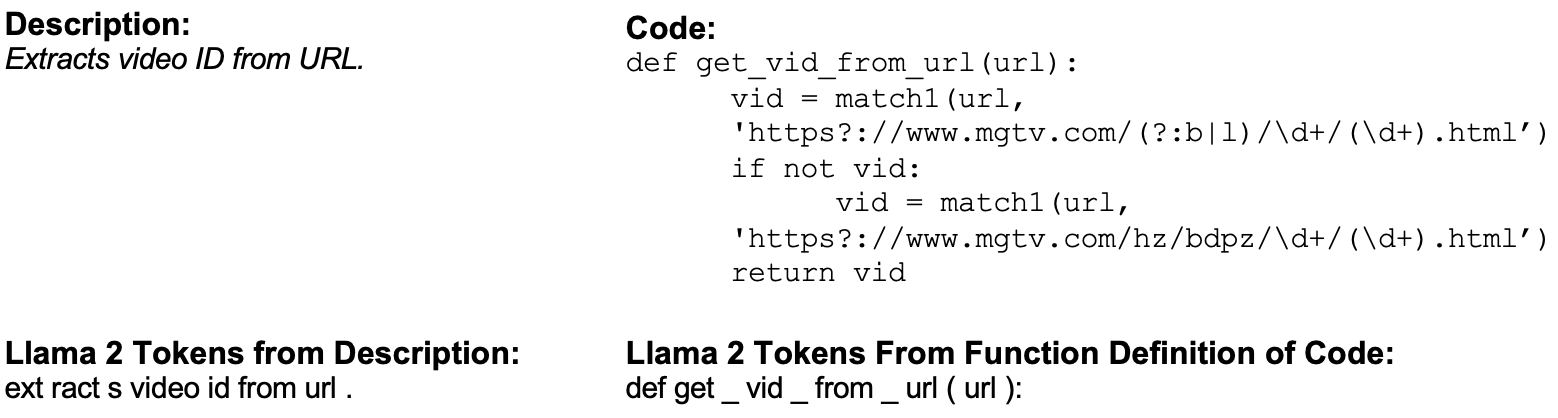}
  \caption{Subword tokenization (here performed by the Llama 2 tokenizer on the first line of the code and the description) exposes valuable information about which tokens are present in the description.}
  \label{fig:llama_token_example}
\end{figure}

Although current LLMs differ in important details (size, amount of training data, pre-training regime, fine-tuning, etc.), they use a form of \textbf{subword tokenization}~\cite{kudo-richardson-2018-sentencepiece,sennrich-etal-2016-neural}, in which words are split into shorter strings (e.g. extracts $\rightarrow$ ext ract s) and underscores are treated as separators (e.g. from\_url $\rightarrow$ from \_ url), such that the model's vocabulary consists of subwords rather than whitespace-delineated tokens. Figure \ref{fig:llama_token_example} shows how even just a function definition reveals valuable information about the tokens that are likely to occur in the reference description when it is tokenized by a subword tokenizer (here, Llama 2). 

\subsection{Code Summarization}

\begin{figure}[!t]
\centering
  \includegraphics[width=0.9\linewidth]{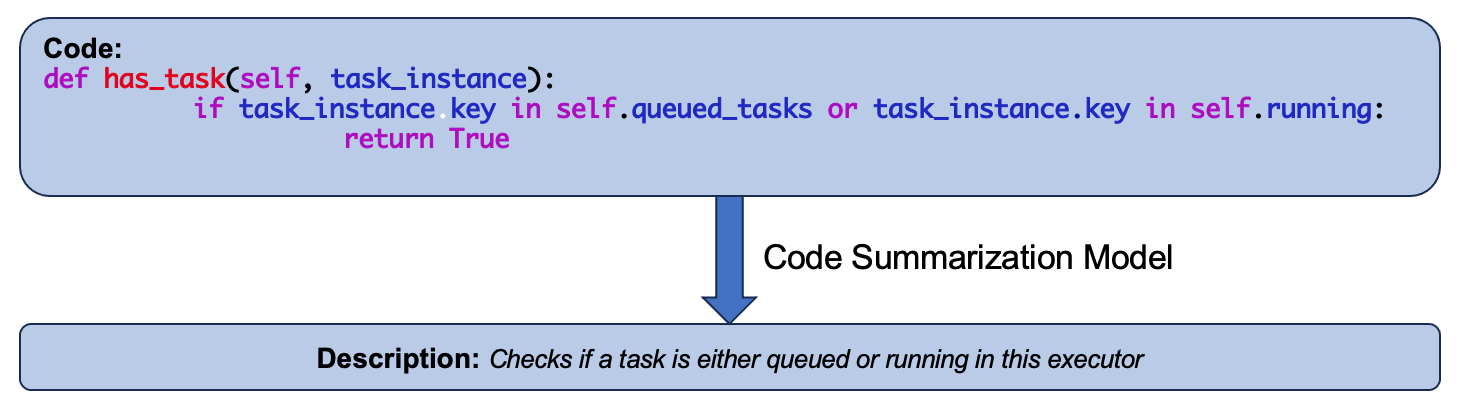}
  \caption{In code summarization, a code snippet (here,  from CodeXGLUE) is given as input to a model that returns an English description.}
  \label{fig:dataset_examples}
\end{figure}

\textbf{Code Summarization} systems translate code snippets into automatically generated English summaries that describe what the code does. Given an input code snippet $\mathit{C}$, the system has to return a description $\mathit{D}$ that accurately describes what that code does. Figure \ref{fig:dataset_examples} shows an example of code summarization being performed.

When the task was first introduced, template-based approaches~\cite{template1} requiring expert domain knowledge were used, followed by information retrieval-based approaches~\cite{ir1,ir2}. In IR-based approaches, the model would extract the most relevant tokens from the code to generate a term-based summary. This did not require any domain knowledge but the models used did not show any deep understanding of the code structure and instead relied on informative function names and comments to generate summaries. This was followed by early neural approaches~\cite{iyer-etal-2016-summarizing}, Encoder-decoder frameworks like~\cite{hu2018deep} and transformer-based approaches, like PLBART~\cite{ahmad-etal-2021-unified}, Structure-induced Transformers~\cite{wu-etal-2021-code}, CoTexT~\cite{phan-etal-2021-cotext}  CodeGPT~\cite{codexglue}, Codex~\cite{chen2021evaluating}, and CodeT5~\cite{wang-etal-2021-codet5}. Recently, these models have been surpassed by large language models using decoder-only architectures like PaLM 2~\cite{anil2023palm}, GPT-4~\cite{openai2023gpt4}, and Llama 2~\cite{touvron2023llama}.

Benchmark datasets for this task include TL-CodeSum~\cite{tlcodesumm}, Funcom~\cite{funcom}, CodeSearchNet~\cite{husain2019codesearchnet} and more recently, CodeXGlUE~\cite{codexglue}.

\paragraph{Dataset:}\textbf{CodeXGLUE}~\cite{codexglue} is a standard benchmark for several code-NL tasks including code summarization. It is a filtered subset of \textbf{CodeSearchNet}~\cite{husain2019codesearchnet}, consisting of code snippets paired with English descriptions that are scraped from public open-source GitHub repositories for Go, Java, JavaScript, PHP, Python, and Ruby.  The description paired with each code snippet is the first paragraph of the repository. Examples were filtered out from this dataset if their code could not be parsed into an abstract syntax tree, or if their descriptions were not in English, contained special strings like "http://", were empty, or not between 3 and 256 tokens in length. Figure \ref{fig:dataset_examples} shows an example from the dataset being used for code summarization. In our experiments, we use the Python examples from CodeXGLUE, which contain 251,820 training, 13,914 dev, and 14,918 testing data points.

\paragraph{Metric:} BLEU~\cite{papineni-etal-2002-bleu} is computed by matching \textit{n}-grams between generated summaries and human written summaries. A perfect match between the two summaries would give a score of 100\%, and a perfect mismatch would give a score of zero. BLEU-4 is the standard metric for evaluating code summarization~\cite{eval_code_summ}. In this case, \textit{n}-grams are matched up to $\mathit{n}=4$. This matching is done cumulatively: the BLEU scores for \textit{n}-grams for values of \textit{n} = 1 through 4 are computed, and then the mean of these four scores is computed to get the final score.

It was found that different implementations of BLEU-4 lead to different results~\cite{eval_code_summ}. Out of these variants, the sentence BLEU metric based on NLTK~\cite{bird2009natural} smoothing method 4 had the highest correlation with human evaluation. The scores we report are computed by the implementation of this method in NLTK 3.8.1.

\section{How well do LLMs perform on Code Summarization?}
\label{sec:our_approach}

We now show how some recent LLMs perform on the standard code summarization dataset, CodeXGLUE.  In this section, we study the extent to which the generated English summaries contain tokens that are copied from the subword-tokenized code. Following this, we examine how prevalent this token overlap is across the entire dataset. We further investigate whether the models also follow the trend of generating summaries with high token overlap with the input code and whether the BLEU scores of the generations are impacted by this overlap. Finally, we inspect which types of tokens are more likely to be involved in this overlap, since the tokens present in function names give away more information about the function than the tokens in the body of the function.

After determining the prevalence of high token overlap in the dataset and its impact on generation performance, we examine how much LLMs prioritize function names over the code structure and its control flow. We also evaluate how they perform when given code that has all the control structure removed, and in an extreme case if we only give the function definition as input. This will also help us answer our first two research questions about how much models leverage function names over the body of the code and how much models care about the code syntax and structure. Additionally, studying the relative performance of multiple LLMs of different parameter counts on these different types of input will give us a deeper insight into what they understand about code. While there is prior work showing how transformations like masking function names make code summarization harder~\cite{sontakke2022code}, our work goes further and performs a deeper qualitative analysis to see what causes the model to perform well in the first place.

\subsection{Experimental Setup}
To examine a wide range of model sizes, we analyze CodeT5 (220M parameters), PaLM 2 (340B parameters), and Llama 2 (with 7B and 70B parameters): 

\textbf{CodeT5:} CodeT5~\cite{wang-etal-2021-codet5} is a pre-trained encoder-decoder model that uses objectives during pre-training like Identifier Tagging (the model has to predict whether each token in the input is an identifier or not) which make it more suitable for understanding code. It is pre-trained on CodeSearchNet. It is adapted to multiple downstream tasks including code summarization and code generation through multi-task learning.

\textbf{PaLM 2:} PaLM 2~\cite{anil2023palm} is an LLM  made by Google AI for reasoning tasks, question answering, classification, translation, and code summarization. It was pre-trained on a large corpus of parallel multilingual text. As it is proficient in sequence-to-sequence tasks, we can make it perform code summarization with the right prompt.

\textbf{Llama 2:} Llama 2~\cite{touvron2023llama} is an LLM designed by Meta AI that can also be used for sequence-to-sequence tasks, including an instruct model that can be asked to perform any task with a prompt. It comes in three sizes - 7B, 13B and 70B.

While CodeT5 is fine-tuned on this dataset, we use the other models PaLM 2 and Llama 2 in inference-only mode with few-shot prompting. We use the following prompt before each input code snippet - \textit{"Pretend that you are a programmer writing Python functions. For a given Python function you have to generate a short documentation describing what the function does."}, along with ten examples from the dataset to show what kind of description we are looking for.

\subsection{Overall Performance}
To analyze these models, we first evaluate them on code summarization on the standard CodeXGLUE dataset. Table \ref{tab:bleu_first_row} shows that the BLEU-4 scores of the models discussed above range from 17.66 for the 220M-parameter model CodeT5 to 22.41 for the 70B-parameter Llama 2. 
The much smaller 7b parameter version of Llama 2 performs similarly at 22.36, while PalM 2 (340B parameters) achieves a score of 19.23,   between CodeT5 and Llama 2.

\begin{table}[tb]
\centering
\begin{scriptsize}
\begin{tabular}{@{}lllll@{}}
\toprule
\textbf{Dataset} & \textbf{CodeT5}           & \textbf{PaLM 2}           & \textbf{Llama 2 (7b)}     & \textbf{Llama 2 (70b)}    \\ \midrule
CodeXGLUE        & \multicolumn{1}{r}{17.66} & \multicolumn{1}{r}{19.23} & \multicolumn{1}{r}{22.36} & \multicolumn{1}{r}{22.41} \\ \bottomrule
\end{tabular}
\end{scriptsize}
\caption{BLEU-4 across all models on CodeXGLUE}
\label{tab:bleu_first_row}
\end{table}

\subsection{Exploring if LLMs are copying tokens from code to description}

We now analyze if these models are copying tokens from the code when generating their summaries. While copying in itself would not be a design flaw, if there is an instance of widespread copying, this would indicate that it is possible to perform well in this task without showing much understanding of the semantics of the code. We define our own metric called $\mathit{p_{copy}}$, which is the percentage of tokens (as generated by the corresponding model's tokenizer) in the description that was also present in the code.

\begin{figure}[h]
\centering
\begin{small}
  \begin{subfigure}[b]{\columnwidth}
  \centering
    \includegraphics[width=\linewidth]{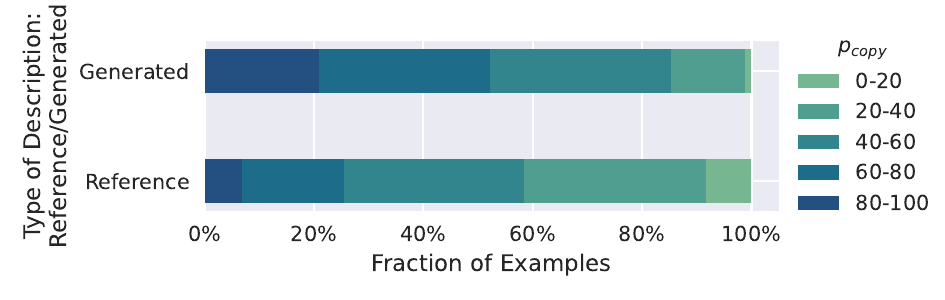}
    \caption{CodeT5}
  \end{subfigure}
  \begin{subfigure}[b]{\columnwidth}
  \centering
    {\includegraphics[width=\linewidth]{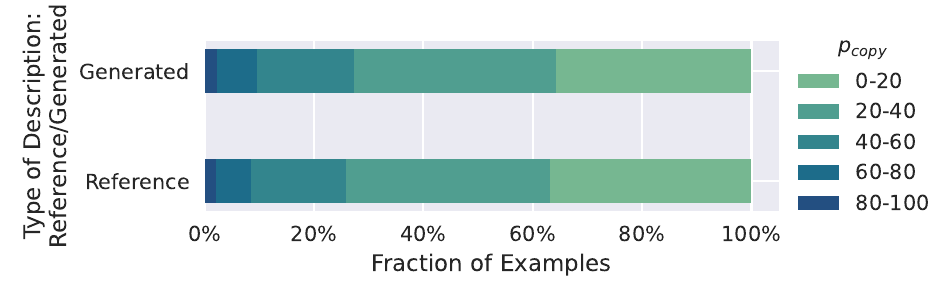}}
    \caption{Llama 2 (70b)}
  \end{subfigure}
  \end{small}
  \caption{Distribution of $\mathit{p_{copy}}$ in the dataset. Generated descriptions have much higher $\mathit{p_{copy}}$ than Reference descriptions, and Llama 2's has fewer issues with copying tokens than CodeT5.}
   \label{fig:stacked_hist}
\end{figure}

Figure \ref{fig:stacked_hist} shows the distribution of $\mathit{p_{copy}}$ in the reference descriptions written by human developers (bottom) and in the summaries produced by our two models (top), CodeT5 and Llama 2. Although copying is present in the human descriptions in both models (allowing the model to pick up on it during training), the models rely on it to a greater extent, since $\mathit{p_{copy}}$ skews higher in the generated summaries. CodeT5 generates descriptions with much higher copying than Llama 2, but we see that even in the reference descriptions CodeT5 tokenizer gives much higher overlap than Llama 2. Therefore, it is not just the model but also the tokenizer that greatly influences the extent to which this copying strategy is employed.

\subsection{Is copying tokens a viable strategy for summarization?}

\begin{figure}[t]
\centering
\begin{small}
  \begin{subfigure}[b]{0.4\textwidth}
  \centering
    \includegraphics[width=\linewidth]{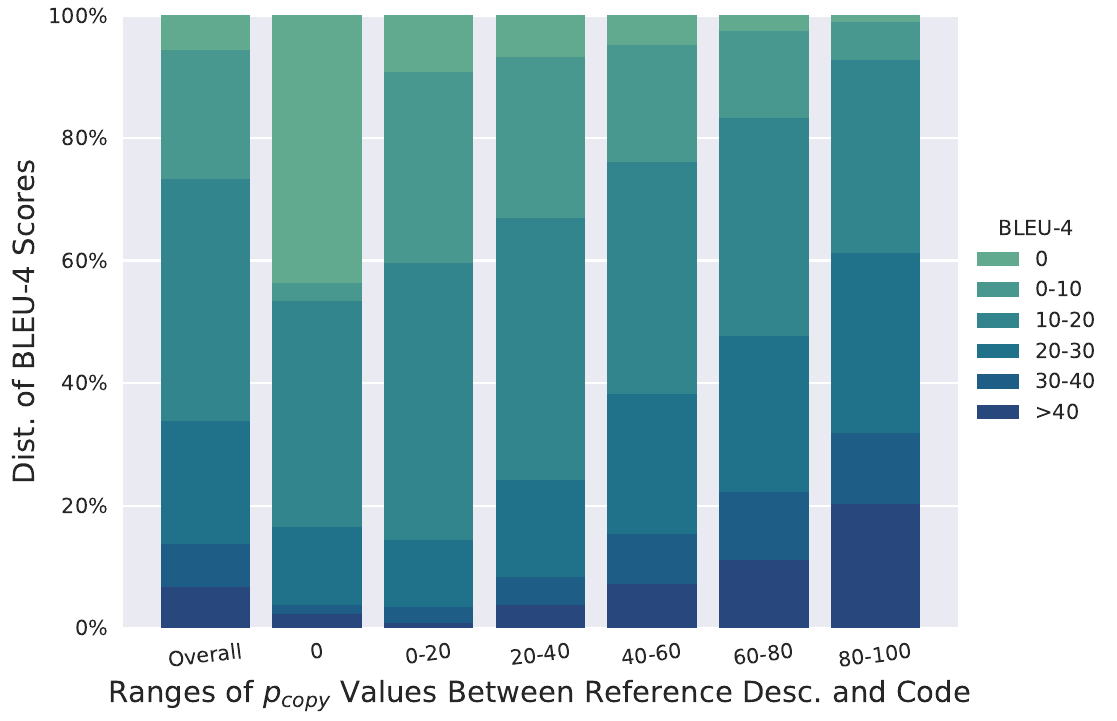}
    \caption{CodeT5}
    \label{fig:pcopy_codet5}
  \end{subfigure}
  \begin{subfigure}[b]{0.4\textwidth}
  \centering
    {\includegraphics[width=\linewidth]{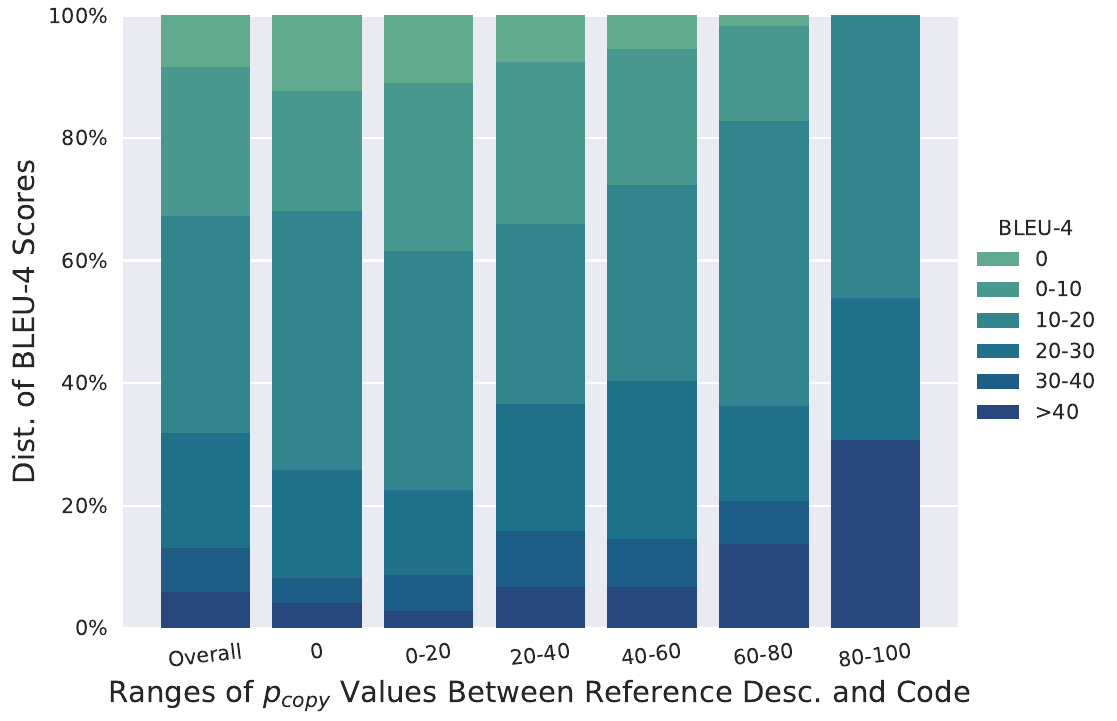}}
    \caption{Llama 2 (70b)}
    \label{fig:pcopy_llama2}
  \end{subfigure}
  \end{small}
  \caption{Distributions of BLEU-4 on all test examples and in different $\mathit{p_{copy}}$ buckets. In general, higher $\mathit{p_{copy}}$ leads to a higher BLEU-4 score.}
   \label{fig:stacked_bleu_to}
\end{figure}

We see that the models learn to generate summaries by copying tokens from the code, but how effective is that strategy? 
For examples with high $\mathit{p_{copy}}$ in the reference description, a copying strategy should yield higher BLEU scores than for examples with lower $\mathit{p_{copy}}$. To see if this is the case, we split the test set into buckets based on the reference descriptions' $\mathit{p_{copy}}$, and plot the distribution of the model's BLEU-4 scores in each bucket (Figure \ref{fig:stacked_bleu_to}). Figure \ref{fig:pcopy_codet5} shows the results on the smallest model we have tested, CodeT5, and Figure \ref{fig:pcopy_llama2} shows the results on Llama 2 (70b).

The first bar of the two plots in Figure \ref{fig:stacked_bleu_to} shows the overall distribution of BLEU-4 scores. The subsequent bars show the distribution of scores for a certain $\mathit{p_{copy}}$ range. For example, the second bar shows the scores for examples where 0 to 10\% of the tokens in the reference description are present in the code. We see that in general as $\mathit{p_{copy}}$ increases, the BLEU-4 scores also increase. However, this correlation is less pronounced for Llama 2. This shows that both models tend to fall back on copying tokens from code to the description as a viable strategy, with the most difficult examples in the dataset being the ones where $\mathit{p_{copy}}$ is zero, as in this bucket, half have a zero BLEU score.

\subsection{Which tokens are copied?}
\label{sec:which_tokens}
We see that copying tokens from the code when generating descriptions can be a viable strategy. However, we need to explore which types of tokens are most likely to be copied. If it is mostly the function names that are being copied, that suggests the model relies most on the function definition and largely ignores the semantics of the code. We also want to see what kind of tokens get copied from the code in the reference descriptions in the dataset.

\begin{figure}[t]
\centering
\begin{small}
  \begin{subfigure}[b]{0.4\textwidth}
  \centering
    \includegraphics[width=\linewidth]{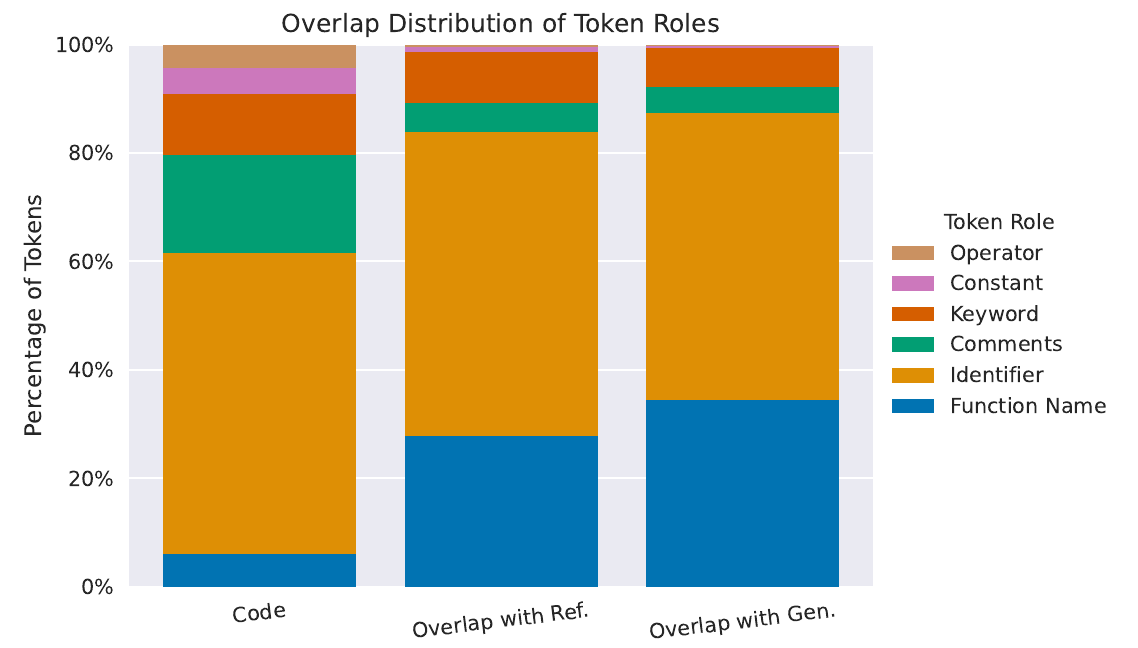}
    \caption{CodeT5}
    \label{fig:dist_tokens_codet5}
  \end{subfigure}
  \begin{subfigure}[b]{0.4\textwidth}
  \centering
    \includegraphics[width=\linewidth]{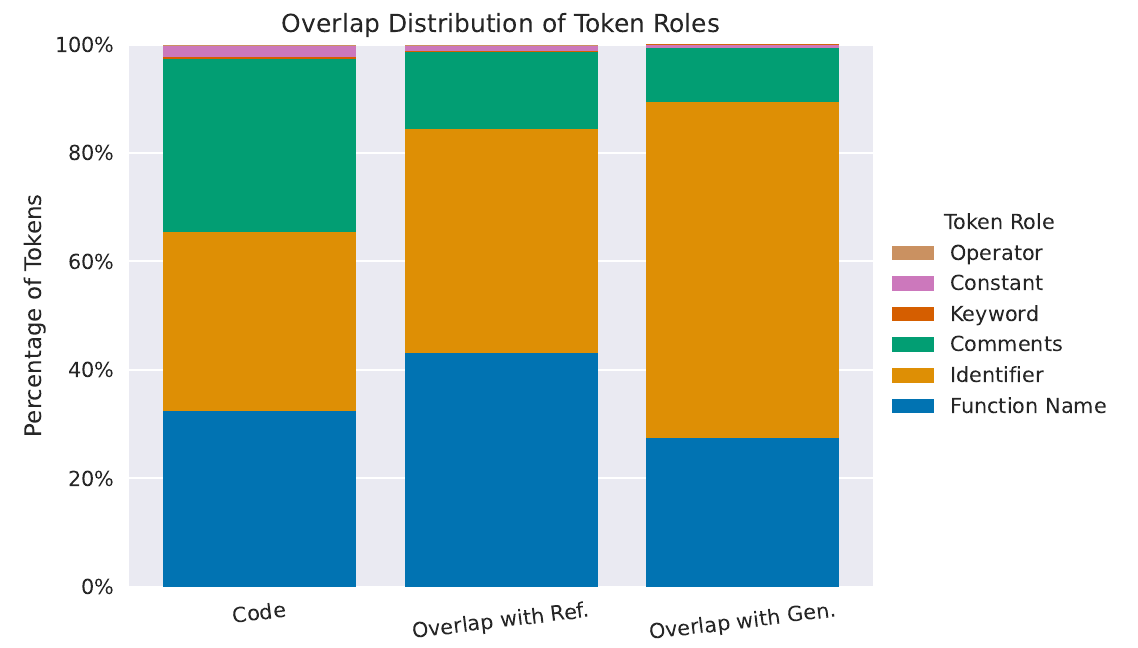} 
    \caption{Llama 2 (70b)}
    \label{fig:dist_tokens_llama2}
  \end{subfigure}
  \end{small}
  \caption{Distributions of token types in the overall CodeXGLUE code, and among the tokens that are copied to the reference and generated descriptions, according to the tokenizer and model of Code T5 and Llama 2 (70b)}
   \label{fig:dist_tokens_overlap}
\end{figure}

In Figure \ref{fig:dist_tokens_overlap} we see the distribution of different token types like function names, identifiers, and comments for both CodeT5 and Llama 2 (70b). One main reason for the differences in their distributions, even in the reference descriptions, is the fact that they use different tokenizers. For example, the function names form a small part of the code snippets after tokenization by CodeT5 as seen in the first column in Figure \ref{fig:dist_tokens_codet5}. On the other hand, the Llama 2 tokenizer (Figure \ref{fig:dist_tokens_llama2}) allows for function names to be a much bigger part of the code. This pattern also holds true for the reference description in the dataset, with function names being more represented by the Llama 2 tokenizer. However, when we look at the generated descriptions, we see that CodeT5-generated summaries have a much higher presence of function names than Llama 2. This is probably due to fine-tuning, and CodeT5 learning that the optimal strategy for generating good summaries is to not only copy tokens from the code but to copy tokens from the function definition of the code. On the other hand, Llama 2 seems to prefer to rely more on identifiers, which can also contain useful information about the code semantics. Llama 2 also seems to better understand that the keywords in the code, while important to understand the semantics, are not supposed to be in the generated description.

\section{How do Code Transformations Affect Performance?}
\label{sec:translation}

We saw that the performance of LLMs on code summarization benefits from the high token overlap between the code and the reference description, and this overlap occurs primarily due to informative function names. To see how well they work when this information is unavailable, and also how strongly they rely on the internal structure of the code instead, we make several transformations to the code in the dataset. We then evaluate the performance of multiple LLMs under these scenarios.

In each transformation, we modify one aspect of the dataset used for training and evaluation of the classifier to understand what effect it has on performance. For example, the change in performance when we remove function names will show us how important function names are to the model's understanding of the code.

\subsection{Code Transformations}
\label{sec:variants}
\begin{figure}[h!]
\centering
  \begin{subfigure}[b]{\columnwidth}
  \centering
    \includegraphics[width=\linewidth]{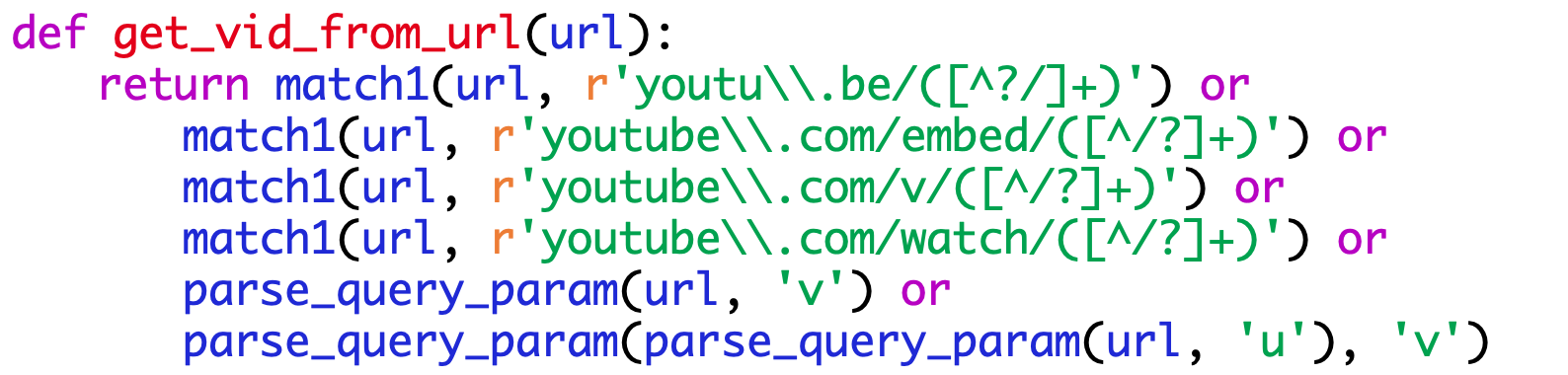}
    \caption{Original Function Names}
    \label{fig:orig_func}
  \end{subfigure}
  \begin{subfigure}[b]{\columnwidth}
  \centering
    {\includegraphics[width=\linewidth]{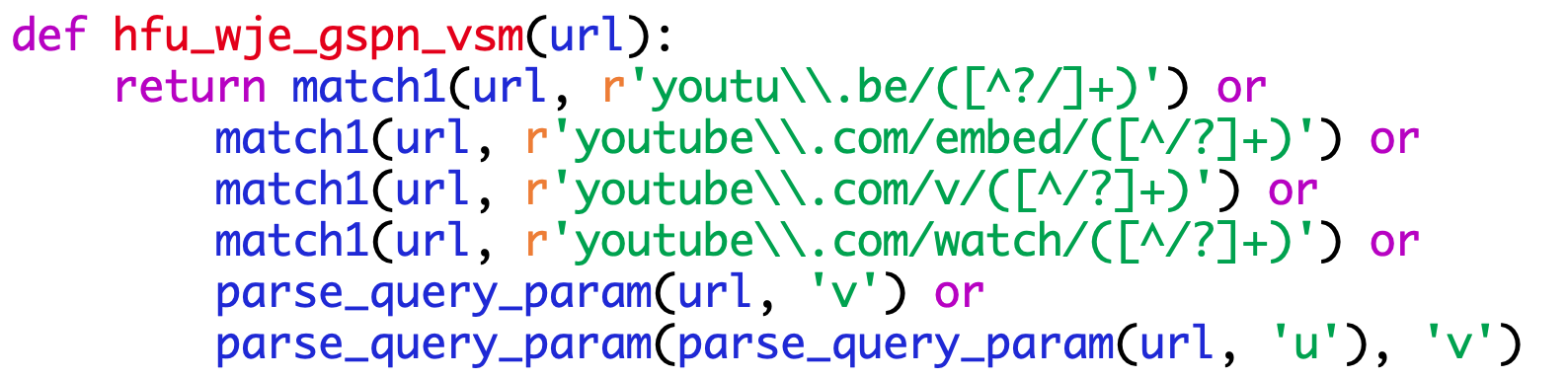}}
    \caption{Obfuscated Function Names}
    \label{fig:obfuscate_func}
  \end{subfigure}
  \begin{subfigure}[b]{\columnwidth}
  \centering
    {\includegraphics[width=\linewidth]{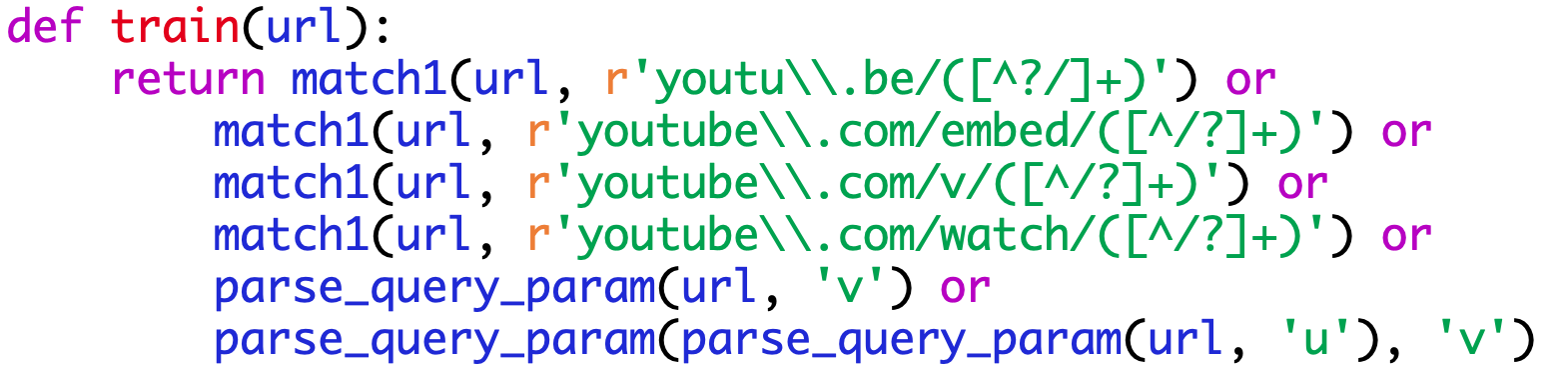}}
    \caption{Adversarial Function Names}
    \label{fig:adv_variant}
  \end{subfigure}
  \begin{subfigure}[b]{\columnwidth}
  \centering
    {\includegraphics[width=\linewidth]{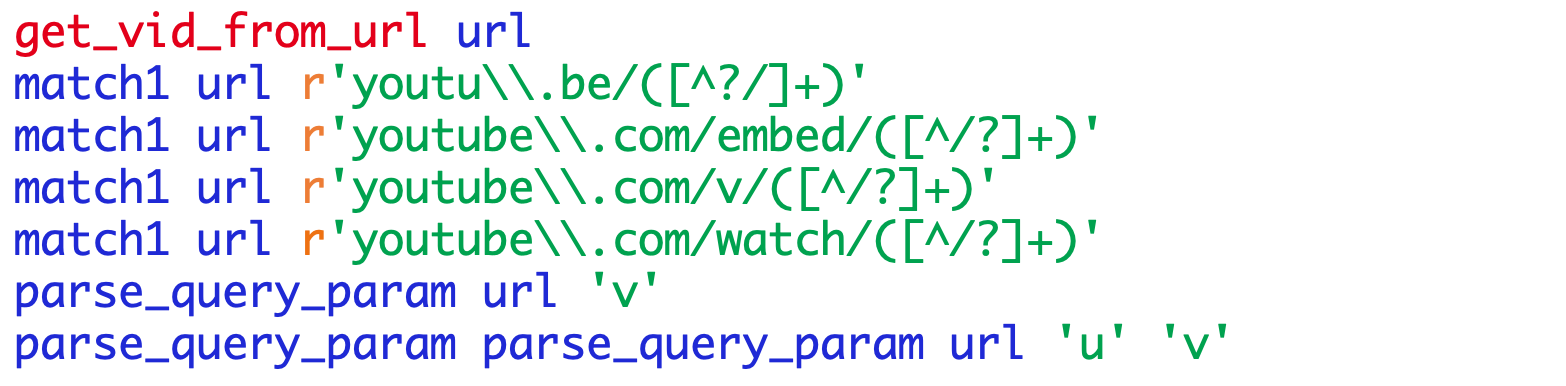}}
    \caption{No Code Structure}
    \label{fig:remove_struct}
  \end{subfigure}
  \begin{subfigure}[b]{\columnwidth}
  \centering
    {\includegraphics[width=\linewidth]{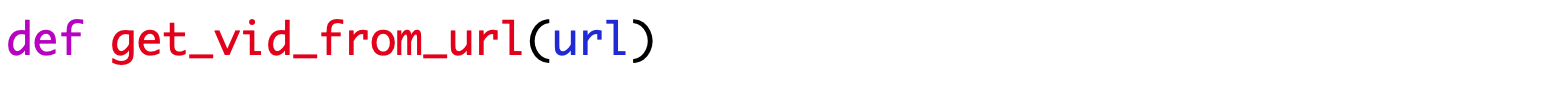}}
    \caption{No Function Body}
    \label{fig:only_func}
  \end{subfigure}
  \caption{Different transformations of a Python function in CodeXGLUE}
   \label{fig:transformation_examples}
\end{figure}

Each of our four variants applies one transformation to the code to remove some information. The corresponding drop in performance indicates the importance of that information for the task. Figure \ref{fig:transformation_examples} shows how an original snippet (\ref{fig:orig_func}), is affected by each transformation. We remove comments from the 32.5\% of snippets that have them so that the models are forced to only look at the code.

\textbf{Original Function Names:} This is the unmodified code from the dataset (Figure \ref{fig:orig_func}).

\textbf{Obfuscated Function names:} Function names often have a higher token overlap with the query than the rest of the code. We obfuscate them by replacing each character with the next character in the alphabet (‘a’ by ‘b’, ‘b’  by ‘c’ etc.). This forces the model to focus on other cues, like comments, variable names, or the actual structure of the code like for-loops and if-statements (Figure \ref{fig:obfuscate_func}). 

\textbf{Adversarial Function Names:} We replace the original function name with the name of another function. Unlike obfuscation, this may mislead the model. Performance on this variant will tell us how well the model works when the function name is at odds with the actual operations performed in the body of the code (Figure \ref{fig:adv_variant}).

\textbf{No Code Structure:} We remove keywords (\texttt{if}, \texttt{return} etc.), operators (\texttt{not}, \texttt{>}, \texttt{+}), and delimiters (\texttt{\(,,,\), etc.}), removing any information about the underlying logic of the program while keeping the rest of the code intact (Figure \ref{fig:remove_struct}).

\textbf{No Function Body:} We remove the entire body of the code and leave only the function definition. Here, we will observe how well the model performs when it only has the function name and its arguments available (Figure \ref{fig:only_func}).

Examples in these variants will show us what part of the code LLMs tend to prioritize when given an input code snippet to analyze.

After computing the performance across multiple models and datasets, we also investigate how the models' performance is affected by the high token overlap between the code and the reference descriptions. Since this overlap can be exploited by copying tokens from the code to generate descriptions that look correct even without understanding the underlying semantics of the code, we explore how often this occurs. We also look at which parts of the code the models tend to copy tokens from most often i.e. function names versus variable names. Finally, we also looked at how our perception of the performance of these models is affected by our choice of similarity-based metrics which compare the descriptions to a human-generated docstring, since similarity alone may not be a perfect judge of whether a generated description is good or not.

\subsection{Performance on Transformed Code}

\begin{table*}
\centering
\small{\begin{tabular}{@{}lrrrrr@{}}
\toprule
\textbf{Variant}           & \multicolumn{1}{l}{\textbf{CodeT5}} & \multicolumn{1}{l}{\textbf{CodeT5 (FT)}} & \multicolumn{1}{l}{\textbf{PaLM 2}} & \multicolumn{1}{l}{\textbf{Llama 2 (7b)}} & \multicolumn{1}{l}{\textbf{Llama 2 (70b)}} \\ \midrule
\textbf{Original Function Names}    & \textbf{17.66}                               & \textbf{17.66}                                    & \textbf{19.23}                               & \textbf{22.36}                                     & \textbf{22.41}                                      \\
Obfuscated Function Names  & 11.59                               & 14.80                                    & 18.72                               & 20.09                                     & 21.25                                      \\
Adversarial Function Names & 11.34                               & 13.12                                    & 15.69                               & 19.53                                     & 21.23                                      \\
No Code Structure          & 13.96                               & 16.57                                    & 11.92                               & 15.11                                     & 18.42                                      \\
No Function Body           & 13.94                               & 15.27                                    & 11.90                               & 14.93                                     & 18.16                                      \\ \bottomrule
\end{tabular}}
\caption{BLEU-4 across all models and variants}
\label{tab:bleu_all_models}
\end{table*}

\begin{table*}
\centering
\small{\begin{tabular}{@{}lrrrrr@{}}
\toprule
\textbf{Variant}           & \multicolumn{1}{l}{\textbf{CodeT5}} & \multicolumn{1}{l}{\textbf{CodeT5 (FT)}} & \multicolumn{1}{l}{\textbf{PaLM 2}} & \multicolumn{1}{l}{\textbf{Llama 2 (7b)}} & \multicolumn{1}{l}{\textbf{Llama 2 (70b)}} \\ \midrule
\textbf{Original Function Names}    & \textbf{83.95}                               & \textbf{83.95}                                    & \textbf{84.26}                               & \textbf{84.34}                                     & \textbf{86.95}                                      \\
Obfuscated Function Names  & 78.57                               & 81.25                                    & 82.28                               & 82.52                                     & 84.41                                      \\
Adversarial Function Names & 78.53                               & 80.01                                    & 80.81                               & 80.84                                     & 84.30                                      \\
No Code Structure          & 81.32                               & 82.86                                    & 79.36                               & 79.66                                     & 83.25                                      \\
No Function Body           & 81.14                               & 82.40                                    & 79.14                               & 79.72                                     & 83.14                                      \\ \bottomrule
\end{tabular}}
\caption{BERTScores across all models and variants}
\label{tab:bertscore_all_models}
\end{table*}

Table \ref{tab:bleu_all_models} shows the BLEU scores of CodeT5, PaLM 2, and Llama 2 (with 7b and 70b parameters) on the different variants, 

The first CodeT5 column shows the scores when the model is only fine-tuned on examples from the original dataset, whereas the second column (FT) shows the scores when the model is fine-tuned on the transformed examples so it knows what kind of code to expect as input.

\begin{figure*}[h!]
\centering
  \includegraphics[width=0.9\textwidth]{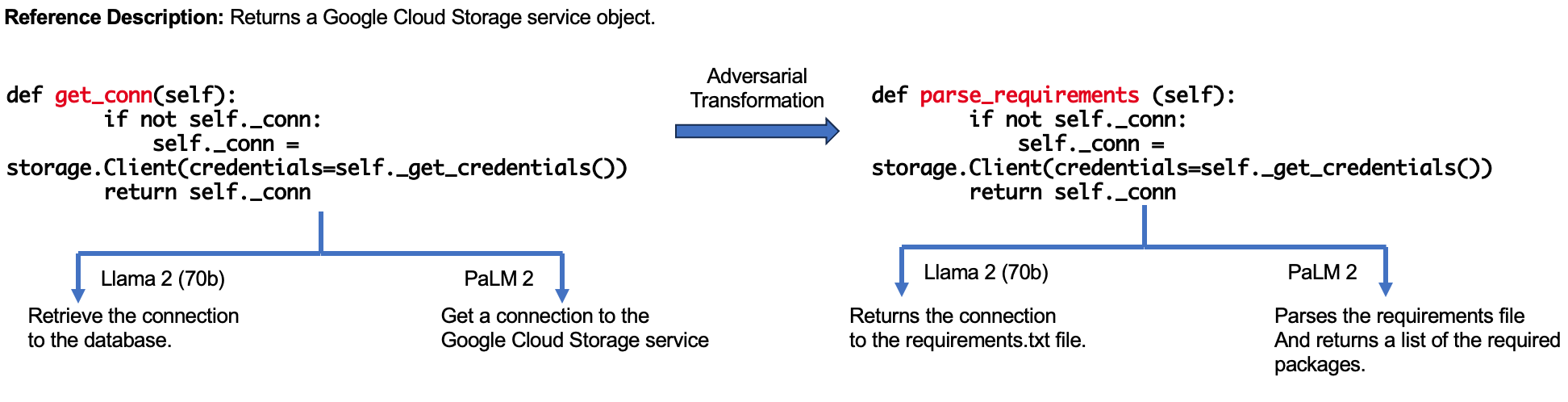}
  \caption{Examples of Llama 2 (70b) and PaLM 2 generations when given code from Adversarial Function Names. We see that even these larger models occasionally falter when given misleading function names. These are in cases where they gave acceptable responses when we made no perturbations in the code}
  \label{fig:adv_gen}
\end{figure*}

\paragraph{Fine-tuning on transformed examples helps CodeT5.} On the transformed variants, the performance of CodeT5 improves significantly after fine-tuning on examples from those variants, especially in Obfuscated Function Names. Not all variants have the same improvement, though. In Adversarial Function Names, the improvement is much smaller, suggesting that the presence of incorrect function names still misleads the model.

\paragraph{Scores drop for the transformed variants.} We see that there is a drop in Obfuscated Function Names and an even bigger drop in Adversarial Function Names, suggesting that models struggle when the function name is hidden and are misled when given an incorrect function name, showing how important function names are to the models' understanding of the code semantics. However, the drop from Obfuscated to Adversarial is much higher for CodeT5 and PaLM 2 than Llama 2, showing that these models are more dependent on the function names. Section \ref{sec:which_tokens} will show that this may be explained by the fact that CodeT5 is much more dependent on the function names than Llama 2.

PaLM 2 and Llama 2 perform worse in No Function Body and No Code Structure compared to the other variants, whereas CodeT5 performs better. This shows that CodeT5 does not mind when given code that is incomplete or syntactically incorrect, unlike the larger models. However, despite performing well on these particular variants, CodeT5 is worse overall than Llama 2, showing that fine-tuning for higher performance on one narrow metric may not always give the best result.

\paragraph{CodeT5 is better at variants that have invalid code.} No Function Body and No Code Structure have incomplete and syntactically incorrect code respectively. CodeT5 performs better at these variants than the ones where we just modify the function names. However, the other models perform much worse here. This shows that the larger variants care more about code syntax whereas a smaller model like CodeT5 looks mostly at function names. In fact, the largest model in our experiments, PaLM 2, performs the worst at these variants.

\paragraph{All models get misled by Adversarial Function Names.} While the bigger models like Llama 2 and PaLM 2 perform better on this variant, they are still prone to be misled. We have shown some examples where it is obvious that the function name is wrong but the models still prioritize that over the semantics in the code. Figure \ref{fig:adv_gen} shows an example where both PaLM 2 and Llama 2 (70b) falter when the function name is changed.

\section{Investigating An Alternative Metric}
\label{sec:low_variance_bertscore}

Although BLEU is a commonly used and well-established metric, it is very strict because it captures only exact matches between n-grams. To address this, a number of metrics have been recently developed that instead compute the similarity of token embeddings returned by a neural model. We investigate whether using such a metric changes the main findings from our experiments. We consider BERTScore, which uses BERT~\cite{devlin-etal-2019-bert} as the neural model.

\textbf{BERTScore}~\cite{bertscore} is a precision/recall-inspired metric that uses a BERT model to compute token similarities between two summaries. To compute BERTScore, both summaries are given as input to the BERT model (we use the base uncased model of DistilBERT~\cite{sanh2020distilbert}) to get their token embeddings (say $x$ for reference and $\hat{x}$ for generated). 
Then recall ($\mathit{R}$)  and precision ($\mathit{P}$) over these embedding sequences are computed by considering the maximal similarity of each reference (candidate) token to any candidate (reference) token: 
\begin{align*}
R &= \frac{1}{|x|}\sum_{x_i \in x} \max_{\hat{x_j} \in \hat{x}} x_i^T\hat{x_j} & P &= \frac{1}{|x|}\sum_{\hat{x_j} \in \hat{x}} \max_{x_i \in x} x_i^T\hat{x_j}
\end{align*}

The final BERTScore is  the \textit{F1} score (harmonic mean ) of $P$ and $R$:
\begin{equation*}
   \mathrm{BERTScore} = \frac{2\cdot P \cdot R}{P+R}
\end{equation*}

\subsection{Overall Performance (BERTScore)}
Table \ref{tab:bertscore_all_models} shows the BERTScores for each model on the original test set. This correlates very strongly with the results in Table \ref{tab:bleu_all_models}, with a Pearson Correlation of 0.87 and a Spearman's Rank Correlation of 0.86 between the 25 pairs of scores.

\begin{figure}[h!]
\centering
  \includegraphics[width=\columnwidth]{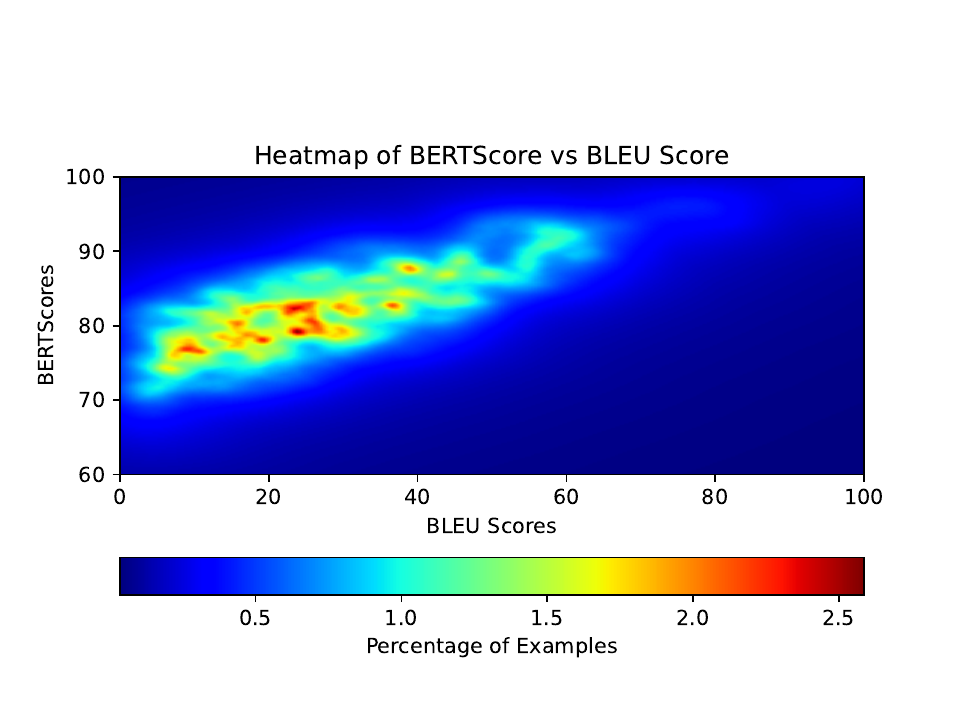}
  \caption{Heatmap of the BERTScore vs the BLEU-4 score for all examples in the test set containing the untransformed code.}
  \label{fig:bleu_dist}
\end{figure}

There is also a high correlation of BLEU and BERTScores between every pair of reference and generated descriptions we have in the original dataset across all models. We get a Pearson Correlation of 0.78 and a Spearman's Rank Correlation of 0.73. The heatmap in Figure \ref{fig:bleu_dist} shows this positive correlation.

However, we observe that BERTScores are overall much higher and closer to each other here than was the case for BLEU scores. This is possibly due to BERTScore being a more forgiving metric, and most generations are assigned a very high BERTScore despite their quality. For example, the minimum BERTScore assigned to an example is 60, whereas there are several cases that have a BLEU score of zero.

Regardless of the metric used, performance generally increases with increasing the number of parameters, except Llama 2 outperforming PaLM 2. However, this increase in performance is not linear, with Llama 2(7b) being quite close to Llama 2(70b). As for variants, all models see a drop in performance in Obfuscated Function Names and an even bigger drop in Adversarial, though the second drop is less in Llama 2. And while Llama 2 and PaLM 2 justifiably underperform when given incomplete or incorrect code in No Function Body and No Code Structure, CodeT5 performance takes a much smaller hit, especially after fine-tuning.

\subsection{Distribution of BERTScores}
\label{sec:bertscore_dist}

\begin{figure}[h!]
\centering
  \includegraphics[width=0.5\textwidth]{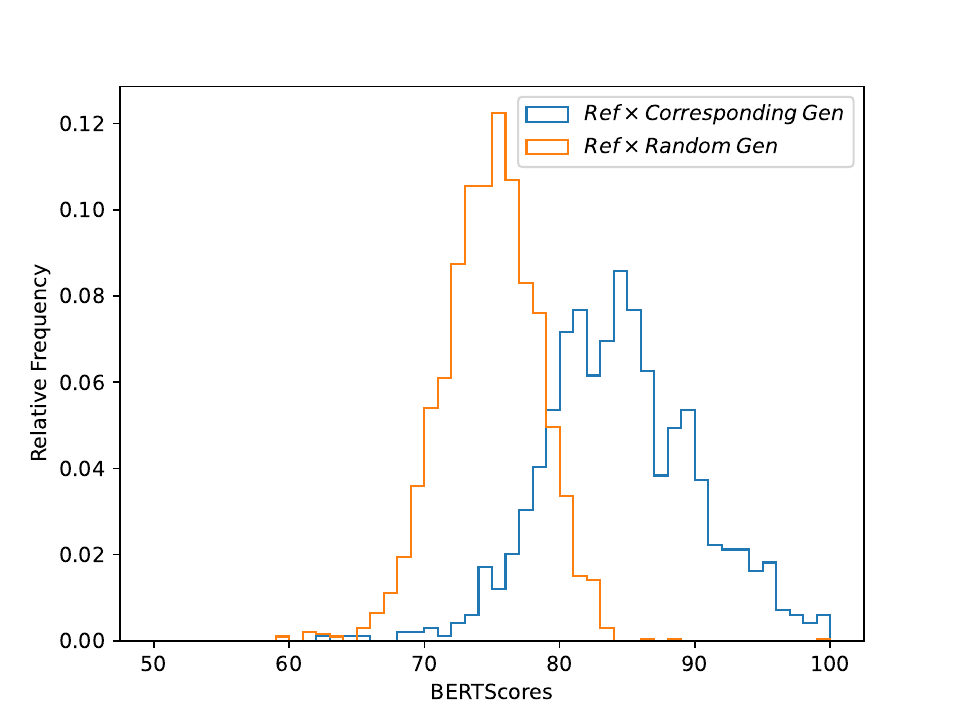}
  \caption{Distribution of BERTscores between reference descriptions and their corresponding generated descriptions, and between reference descriptions and a generated description from another random example across all models}
  \label{fig:bertscore_dist}
\end{figure}

In Figure \ref{fig:bertscore_dist}, we plot the distribution of BERTScores of all the model outputs on the original test data. The orange curve is the distribution of BERTScores between random pairs of reference descriptions and generated descriptions, and the blue curve shows the distribution of BERTScores between the reference descriptions and their corresponding descriptions.

We see that the median BERTScore even between two unrelated descriptions is high, around 75.03. Therefore, even incorrect summaries are assigned very high BERTScores, and the threshold for good summaries is much higher than that. The median BERTScore between the reference descriptions and their corresponding generation as returned by all models is 84.33, which is more than 9 points higher, showing that this metric still shows discernment between related and unrelated descriptions. In Appendix \ref{appendix:dist_bleu}, we see that BLEU scores can discern between correct and incorrect descriptions as well as assign low scores (mostly zero) to randomly sampled generated descriptions.

\begin{figure}[h!]
\centering
  \includegraphics[width=0.5\textwidth]{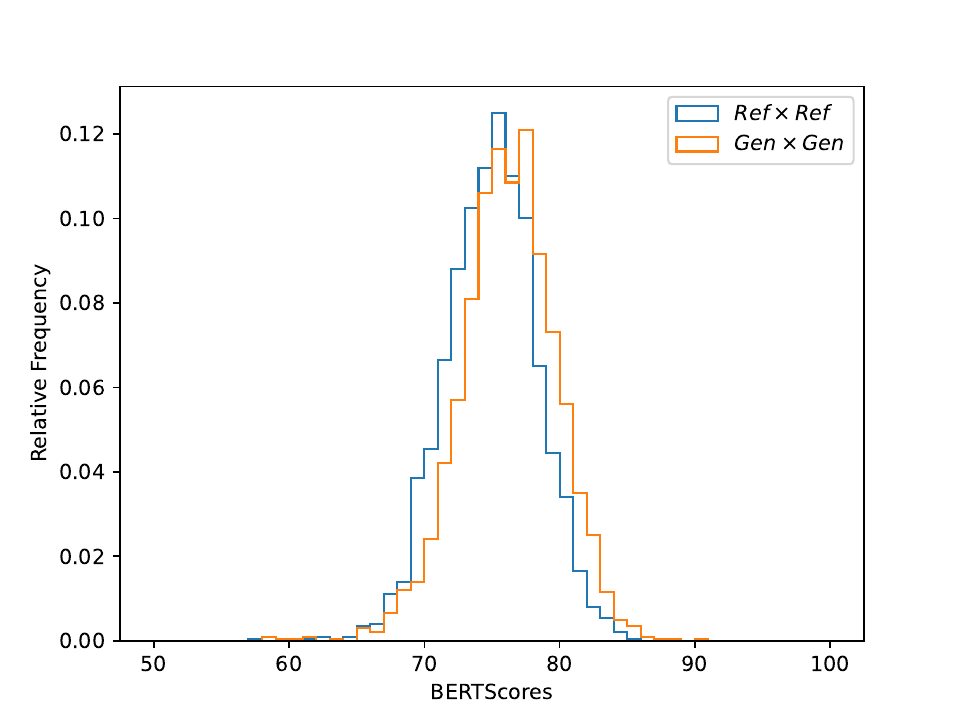}
  \caption{Distribution of BERTScores between two random reference descriptions and two random generated descriptions across all models}
  \label{fig:bertscore_dist_ref_gen}
\end{figure}

Another interesting observation about BERTScores is that the average BERTScore is higher between two random generated descriptions than two random reference descriptions in the dataset, as seen in Figure \ref{fig:bertscore_dist_ref_gen}. This tells us that the model generates descriptions with less diversity than that present in the dataset. In Appendix \ref{appendix:dist_bleu}, we see this is also true for BLEU scores.

\section{Conclusion}

This paper presented a series of experiments to gain a deeper insight into what makes current LLMs effective at code summarization. Section \ref{sec:translation} suggests that LLMs often rely on function names and on shared tokens between the code and the description compared to the code structure to perform well. Relying on token overlap seems to work well because, at least in the standard datasets for these tasks,  code and the corresponding descriptions often have high token overlap. Our experiments establish a clear trend between the token overlap of the code and descriptions and summarization performance.

We expect other LLM-based models, like PLBART~\citep{ahmad-etal-2021-unified}, and CoText~\citep{phan-etal-2021-cotext} to show similar behavior since we believe our results point to a feature of this entire class of models.

The current state of the art also falls short when the description is in a language other than English. \citet{mconala} released a benchmark MCoNaLa, which contains a parallel corpus of code paired with multiple languages, and showed that current code generation models perform poorly for languages like Spanish, Japanese, and Russian. This may be because there are fewer tokens in the description that overlap with the code, so a model cannot learn to take advantage of informative function names and identifier names.

Finally, we also believe that human evaluation should be used in conjunction with building more comprehensive evaluation methodologies for code summarization, in order to measure the accuracy and the usefulness of a generated description, something that similarity-based metrics like BLEU and BERTScore cannot capture. While there has been work in that direction~\cite{eval_code_summ}, the current practice still relies on comparing generated descriptions with a reference instead of measuring their actual usefulness to the user.

\section*{Acknowledgements}
This work is supported by Agriculture and Food Research Initiative (AFRI) grant no. 2020-67021-32799/project accession no.1024178 from the USDA National Institute of Food and Agriculture.

\section*{Bibliographical References}\label{sec:reference}
\bibliographystyle{lrec-coling2024-natbib}
\bibliography{anthology, custom, related}

\begin{thebibliography}{50}
\expandafter\ifx\csname natexlab\endcsname\relax\def\natexlab#1{#1}\fi

\bibitem[{Ahmad et~al.(2021)Ahmad, Chakraborty, Ray, and Chang}]{ahmad-etal-2021-unified}
Wasi Ahmad, Saikat Chakraborty, Baishakhi Ray, and Kai-Wei Chang. 2021.
\newblock \href {https://doi.org/10.18653/v1/2021.naacl-main.211} {Unified pre-training for program understanding and generation}.
\newblock In \emph{Proceedings of the 2021 Conference of the North American Chapter of the Association for Computational Linguistics: Human Language Technologies}, pages 2655--2668, Online. Association for Computational Linguistics.

\bibitem[{Alon et~al.(2019)Alon, Zilberstein, Levy, and Yahav}]{alon2019code2vec}
Uri Alon, Meital Zilberstein, Omer Levy, and Eran Yahav. 2019.
\newblock \href {https://doi.org/10.1145/3290353} {code2vec: learning distributed representations of code}.
\newblock \emph{Proc. ACM Program. Lang.}, 3(POPL).

\bibitem[{Anil et~al.(2023)Anil, Dai, Firat, Johnson, Lepikhin, Passos, Shakeri, Taropa, Bailey, Chen, Chu, Clark, Shafey, Huang, Meier-Hellstern, Mishra, Moreira, Omernick, Robinson, Ruder, Tay, Xiao, Xu, Zhang, Abrego, Ahn, Austin, Barham, Botha, Bradbury, Brahma, Brooks, Catasta, Cheng, Cherry, Choquette-Choo, Chowdhery, Crepy, Dave, Dehghani, Dev, Devlin, Díaz, Du, Dyer, Feinberg, Feng, Fienber, Freitag, Garcia, Gehrmann, Gonzalez, Gur-Ari, Hand, Hashemi, Hou, Howland, Hu, Hui, Hurwitz, Isard, Ittycheriah, Jagielski, Jia, Kenealy, Krikun, Kudugunta, Lan, Lee, Lee, Li, Li, Li, Li, Li, Lim, Lin, Liu, Liu, Maggioni, Mahendru, Maynez, Misra, Moussalem, Nado, Nham, Ni, Nystrom, Parrish, Pellat, Polacek, Polozov, Pope, Qiao, Reif, Richter, Riley, Ros, Roy, Saeta, Samuel, Shelby, Slone, Smilkov, So, Sohn, Tokumine, Valter, Vasudevan, Vodrahalli, Wang, Wang, Wang, Wang, Wieting, Wu, Xu, Xu, Xue, Yin, Yu, Zhang, Zheng, Zheng, Zhou, Zhou, Petrov, and Wu}]{anil2023palm}
Rohan Anil, Andrew~M. Dai, Orhan Firat, Melvin Johnson, Dmitry Lepikhin, Alexandre Passos, Siamak Shakeri, Emanuel Taropa, Paige Bailey, Zhifeng Chen, Eric Chu, Jonathan~H. Clark, Laurent~El Shafey, Yanping Huang, Kathy Meier-Hellstern, Gaurav Mishra, Erica Moreira, Mark Omernick, Kevin Robinson, Sebastian Ruder, Yi~Tay, Kefan Xiao, Yuanzhong Xu, Yujing Zhang, Gustavo~Hernandez Abrego, Junwhan Ahn, Jacob Austin, Paul Barham, Jan Botha, James Bradbury, Siddhartha Brahma, Kevin Brooks, Michele Catasta, Yong Cheng, Colin Cherry, Christopher~A. Choquette-Choo, Aakanksha Chowdhery, Clément Crepy, Shachi Dave, Mostafa Dehghani, Sunipa Dev, Jacob Devlin, Mark Díaz, Nan Du, Ethan Dyer, Vlad Feinberg, Fangxiaoyu Feng, Vlad Fienber, Markus Freitag, Xavier Garcia, Sebastian Gehrmann, Lucas Gonzalez, Guy Gur-Ari, Steven Hand, Hadi Hashemi, Le~Hou, Joshua Howland, Andrea Hu, Jeffrey Hui, Jeremy Hurwitz, Michael Isard, Abe Ittycheriah, Matthew Jagielski, Wenhao Jia, Kathleen Kenealy, Maxim Krikun, Sneha Kudugunta, Chang
  Lan, Katherine Lee, Benjamin Lee, Eric Li, Music Li, Wei Li, YaGuang Li, Jian Li, Hyeontaek Lim, Hanzhao Lin, Zhongtao Liu, Frederick Liu, Marcello Maggioni, Aroma Mahendru, Joshua Maynez, Vedant Misra, Maysam Moussalem, Zachary Nado, John Nham, Eric Ni, Andrew Nystrom, Alicia Parrish, Marie Pellat, Martin Polacek, Alex Polozov, Reiner Pope, Siyuan Qiao, Emily Reif, Bryan Richter, Parker Riley, Alex~Castro Ros, Aurko Roy, Brennan Saeta, Rajkumar Samuel, Renee Shelby, Ambrose Slone, Daniel Smilkov, David~R. So, Daniel Sohn, Simon Tokumine, Dasha Valter, Vijay Vasudevan, Kiran Vodrahalli, Xuezhi Wang, Pidong Wang, Zirui Wang, Tao Wang, John Wieting, Yuhuai Wu, Kelvin Xu, Yunhan Xu, Linting Xue, Pengcheng Yin, Jiahui Yu, Qiao Zhang, Steven Zheng, Ce~Zheng, Weikang Zhou, Denny Zhou, Slav Petrov, and Yonghui Wu. 2023.
\newblock \href {http://arxiv.org/abs/2305.10403} {Palm 2 technical report}.

\bibitem[{Bird et~al.(2009)Bird, Loper, and Klein}]{bird2009natural}
Steven Bird, Edward Loper, and Ewan Klein. 2009.
\newblock \emph{Natural Language Processing with Python}.
\newblock O'Reilly Media Inc.

\bibitem[{Brown et~al.(2020)Brown, Mann, Ryder, Subbiah, Kaplan, Dhariwal, Neelakantan, Shyam, Sastry, Askell, Agarwal, Herbert-Voss, Krueger, Henighan, Child, Ramesh, Ziegler, Wu, Winter, Hesse, Chen, Sigler, Litwin, Gray, Chess, Clark, Berner, McCandlish, Radford, Sutskever, and Amodei}]{NEURIPS2020_1457c0d6}
Tom Brown, Benjamin Mann, Nick Ryder, Melanie Subbiah, Jared~D Kaplan, Prafulla Dhariwal, Arvind Neelakantan, Pranav Shyam, Girish Sastry, Amanda Askell, Sandhini Agarwal, Ariel Herbert-Voss, Gretchen Krueger, Tom Henighan, Rewon Child, Aditya Ramesh, Daniel Ziegler, Jeffrey Wu, Clemens Winter, Chris Hesse, Mark Chen, Eric Sigler, Mateusz Litwin, Scott Gray, Benjamin Chess, Jack Clark, Christopher Berner, Sam McCandlish, Alec Radford, Ilya Sutskever, and Dario Amodei. 2020.
\newblock \href {https://proceedings.neurips.cc/paper/2020/file/1457c0d6bfcb4967418bfb8ac142f64a-Paper.pdf} {Language models are few-shot learners}.
\newblock In \emph{Advances in Neural Information Processing Systems}, volume~33, pages 1877--1901. Curran Associates, Inc.

\bibitem[{Chen et~al.(2021)Chen, Tworek, Jun, Yuan, Pinto, Kaplan, Edwards, Burda, Joseph, Brockman et~al.}]{chen2021evaluating}
Mark Chen, Jerry Tworek, Heewoo Jun, Qiming Yuan, Henrique Ponde de~Oliveira Pinto, Jared Kaplan, Harri Edwards, Yuri Burda, Nicholas Joseph, Greg Brockman, et~al. 2021.
\newblock Evaluating large language models trained on code.
\newblock \emph{arXiv preprint arXiv:2107.03374}.

\bibitem[{Devlin et~al.(2019)Devlin, Chang, Lee, and Toutanova}]{devlin-etal-2019-bert}
Jacob Devlin, Ming-Wei Chang, Kenton Lee, and Kristina Toutanova. 2019.
\newblock \href {https://doi.org/10.18653/v1/N19-1423} {{BERT}: Pre-training of deep bidirectional transformers for language understanding}.
\newblock In \emph{Proceedings of the 2019 Conference of the North {A}merican Chapter of the Association for Computational Linguistics: Human Language Technologies, Volume 1 (Long and Short Papers)}, pages 4171--4186, Minneapolis, Minnesota. Association for Computational Linguistics.

\bibitem[{Eddy et~al.(2013)Eddy, Robinson, Kraft, and Carver}]{ir1}
Brian~P. Eddy, Jeffrey~A. Robinson, Nicholas~A. Kraft, and Jeffrey~C. Carver. 2013.
\newblock \href {https://doi.org/10.1109/ICPC.2013.6613829} {Evaluating source code summarization techniques: Replication and expansion}.
\newblock In \emph{2013 21st International Conference on Program Comprehension (ICPC)}, pages 13--22.

\bibitem[{Feng et~al.(2020)Feng, Guo, Tang, Duan, Feng, Gong, Shou, Qin, Liu, Jiang, and Zhou}]{feng-etal-2020-codebert}
Zhangyin Feng, Daya Guo, Duyu Tang, Nan Duan, Xiaocheng Feng, Ming Gong, Linjun Shou, Bing Qin, Ting Liu, Daxin Jiang, and Ming Zhou. 2020.
\newblock \href {https://doi.org/10.18653/v1/2020.findings-emnlp.139} {{C}ode{BERT}: A pre-trained model for programming and natural languages}.
\newblock In \emph{Findings of the Association for Computational Linguistics: EMNLP 2020}, pages 1536--1547, Online. Association for Computational Linguistics.

\bibitem[{Gu et~al.(2018{\natexlab{a}})Gu, Zhang, and Kim}]{deepcs}
Xiaodong Gu, Hongyu Zhang, and Sunghun Kim. 2018{\natexlab{a}}.
\newblock \href {https://doi.org/10.1145/3180155.3180167} {Deep code search}.
\newblock In \emph{2018 IEEE/ACM 40th International Conference on Software Engineering (ICSE)}, pages 933--944.

\bibitem[{Gu et~al.(2018{\natexlab{b}})Gu, Zhang, and Kim}]{gu2018deep}
Xiaodong Gu, Hongyu Zhang, and Sunghun Kim. 2018{\natexlab{b}}.
\newblock \href {https://doi.org/10.1145/3180155.3180167} {Deep code search}.
\newblock In \emph{2018 IEEE/ACM 40th International Conference on Software Engineering (ICSE)}, pages 933--944.

\bibitem[{Guo et~al.(2020)Guo, Ren, Lu, Feng, Tang, Liu, Zhou, Duan, Svyatkovskiy, Fu, Tufano, Deng, Clement, Drain, Sundaresan, Yin, Jiang, and Zhou}]{graphcodebert}
Daya Guo, Shuo Ren, Shuai Lu, Zhangyin Feng, Duyu Tang, Shujie Liu, Long Zhou, Nan Duan, Alexey Svyatkovskiy, Shengyu Fu, Michele Tufano, Shao~Kun Deng, Colin~B. Clement, Dawn Drain, Neel Sundaresan, Jian Yin, Daxin Jiang, and Ming Zhou. 2020.
\newblock \href {http://arxiv.org/abs/2009.08366} {Graphcodebert: Pre-training code representations with data flow}.
\newblock \emph{CoRR}, abs/2009.08366.

\bibitem[{Haldar et~al.(2020)Haldar, Wu, Xiong, and Hockenmaier}]{haldar-etal-2020-multi}
Rajarshi Haldar, Lingfei Wu, JinJun Xiong, and Julia Hockenmaier. 2020.
\newblock \href {https://doi.org/10.18653/v1/2020.acl-main.758} {A multi-perspective architecture for semantic code search}.
\newblock In \emph{Proceedings of the 58th Annual Meeting of the Association for Computational Linguistics}, pages 8563--8568, Online. Association for Computational Linguistics.

\bibitem[{Hu et~al.(2018{\natexlab{a}})Hu, Li, Xia, Lo, and Jin}]{hu2018deep}
Xing Hu, Ge~Li, Xin Xia, David Lo, and Zhi Jin. 2018{\natexlab{a}}.
\newblock \href {https://doi.org/10.1145/3196321.3196334} {Deep code comment generation}.
\newblock In \emph{Proceedings of the 26th Conference on Program Comprehension}, ICPC '18, page 200–210, New York, NY, USA. Association for Computing Machinery.

\bibitem[{Hu et~al.(2018{\natexlab{b}})Hu, Li, Xia, Lo, Lu, and Jin}]{tlcodesumm}
Xing Hu, Ge~Li, Xin Xia, David Lo, Shuai Lu, and Zhi Jin. 2018{\natexlab{b}}.
\newblock Summarizing source code with transferred api knowledge.
\newblock In \emph{Proceedings of the 27th International Joint Conference on Artificial Intelligence}, IJCAI'18, page 2269–2275. AAAI Press.

\bibitem[{Husain et~al.(2019)Husain, Wu, Gazit, Allamanis, and Brockschmidt}]{husain2019codesearchnet}
Hamel Husain, Ho-Hsiang Wu, Tiferet Gazit, Miltiadis Allamanis, and Marc Brockschmidt. 2019.
\newblock {CodeSearchNet} challenge: Evaluating the state of semantic code search.
\newblock \emph{arXiv preprint arXiv:1909.09436}.

\bibitem[{Iyer et~al.(2016)Iyer, Konstas, Cheung, and Zettlemoyer}]{iyer-etal-2016-summarizing}
Srinivasan Iyer, Ioannis Konstas, Alvin Cheung, and Luke Zettlemoyer. 2016.
\newblock \href {https://doi.org/10.18653/v1/P16-1195} {Summarizing source code using a neural attention model}.
\newblock In \emph{Proceedings of the 54th Annual Meeting of the Association for Computational Linguistics (Volume 1: Long Papers)}, pages 2073--2083, Berlin, Germany. Association for Computational Linguistics.

\bibitem[{Kudo and Richardson(2018)}]{kudo-richardson-2018-sentencepiece}
Taku Kudo and John Richardson. 2018.
\newblock \href {https://doi.org/10.18653/v1/D18-2012} {{S}entence{P}iece: A simple and language independent subword tokenizer and detokenizer for neural text processing}.
\newblock In \emph{Proceedings of the 2018 Conference on Empirical Methods in Natural Language Processing: System Demonstrations}, pages 66--71, Brussels, Belgium. Association for Computational Linguistics.

\bibitem[{LeClair et~al.(2019)LeClair, Jiang, and McMillan}]{funcom}
Alexander LeClair, Siyuan Jiang, and Collin McMillan. 2019.
\newblock \href {https://doi.org/10.1109/ICSE.2019.00087} {A neural model for generating natural language summaries of program subroutines}.
\newblock In \emph{Proceedings of the 41st International Conference on Software Engineering}, ICSE '19, page 795–806. IEEE Press.

\bibitem[{Ling et~al.(2021)Ling, Wu, Wang, Pan, Ma, Xu, Liu, Wu, and Ji}]{ling2021}
Xiang Ling, Lingfei Wu, Saizhuo Wang, Gaoning Pan, Tengfei Ma, Fangli Xu, Alex~X. Liu, Chunming Wu, and Shouling Ji. 2021.
\newblock \href {https://doi.org/10.1145/3447571} {Deep graph matching and searching for semantic code retrieval}.
\newblock \emph{ACM Trans. Knowl. Discov. Data}, 15(5).

\bibitem[{Liu et~al.(2021)Liu, Xie, Ma, Siow, and Liu}]{liu2021graphsearchnet}
Shangqing Liu, Xiaofei Xie, Lei Ma, Jingkai Siow, and Yang Liu. 2021.
\newblock Graphsearchnet: Enhancing gnns via capturing global dependency for semantic code search.
\newblock \emph{arXiv preprint arXiv:2111.02671}.

\bibitem[{Liu et~al.(2019)Liu, Ott, Goyal, Du, Joshi, Chen, Levy, Lewis, Zettlemoyer, and Stoyanov}]{roberta}
Yinhan Liu, Myle Ott, Naman Goyal, Jingfei Du, Mandar Joshi, Danqi Chen, Omer Levy, Mike Lewis, Luke Zettlemoyer, and Veselin Stoyanov. 2019.
\newblock \href {http://arxiv.org/abs/1907.11692} {Roberta: {A} robustly optimized {BERT} pretraining approach}.
\newblock \emph{CoRR}, abs/1907.11692.

\bibitem[{Lu et~al.(2021)Lu, Guo, Ren, Huang, Svyatkovskiy, Blanco, Clement, Drain, Jiang, Tang, Li, Zhou, Shou, Zhou, Tufano, Gong, Zhou, Duan, Sundaresan, Deng, Fu, and Liu}]{codexglue}
Shuai Lu, Daya Guo, Shuo Ren, Junjie Huang, Alexey Svyatkovskiy, Ambrosio Blanco, Colin~B. Clement, Dawn Drain, Daxin Jiang, Duyu Tang, Ge~Li, Lidong Zhou, Linjun Shou, Long Zhou, Michele Tufano, Ming Gong, Ming Zhou, Nan Duan, Neel Sundaresan, Shao~Kun Deng, Shengyu Fu, and Shujie Liu. 2021.
\newblock {CodeXGLUE}: {A} machine learning benchmark dataset for code understanding and generation.
\newblock \emph{CoRR}, abs/2102.04664.

\bibitem[{Miceli~Barone and Sennrich(2017)}]{miceli-barone-sennrich-2017-parallel}
Antonio~Valerio Miceli~Barone and Rico Sennrich. 2017.
\newblock \href {https://aclanthology.org/I17-2053} {A parallel corpus of python functions and documentation strings for automated code documentation and code generation}.
\newblock In \emph{Proceedings of the Eighth International Joint Conference on Natural Language Processing (Volume 2: Short Papers)}, pages 314--319, Taipei, Taiwan. Asian Federation of Natural Language Processing.

\bibitem[{Narayanan and Kapoor(2023)}]{narayanan_kapoor_2023}
Arvind Narayanan and Sayash Kapoor. 2023.
\newblock \href {https://aisnakeoil.substack.com/p/gpt-4-and-professional-benchmarks} {Gpt-4 and professional benchmarks: The wrong answer to the wrong question}.

\bibitem[{Nijkamp et~al.(2023)Nijkamp, Pang, Hayashi, Tu, Wang, Zhou, Savarese, and Xiong}]{Nijkamp2022CG}
Erik Nijkamp, Bo~Pang, Hiroaki Hayashi, Lifu Tu, Huan Wang, Yingbo Zhou, Silvio Savarese, and Caiming Xiong. 2023.
\newblock Codegen: An open large language model for code with multi-turn program synthesis.
\newblock \emph{ICLR}.

\bibitem[{OpenAI(2023)}]{openai2023gpt4}
OpenAI. 2023.
\newblock \href {http://arxiv.org/abs/2303.08774} {Gpt-4 technical report}.

\bibitem[{Papineni et~al.(2002)Papineni, Roukos, Ward, and Zhu}]{papineni-etal-2002-bleu}
Kishore Papineni, Salim Roukos, Todd Ward, and Wei-Jing Zhu. 2002.
\newblock \href {https://doi.org/10.3115/1073083.1073135} {{B}leu: a method for automatic evaluation of machine translation}.
\newblock In \emph{Proceedings of the 40th Annual Meeting of the Association for Computational Linguistics}, pages 311--318, Philadelphia, Pennsylvania, USA. Association for Computational Linguistics.

\bibitem[{Phan et~al.(2021)Phan, Tran, Le, Nguyen, Annibal, Peltekian, and Ye}]{phan-etal-2021-cotext}
Long Phan, Hieu Tran, Daniel Le, Hieu Nguyen, James Annibal, Alec Peltekian, and Yanfang Ye. 2021.
\newblock \href {https://doi.org/10.18653/v1/2021.nlp4prog-1.5} {{C}o{T}ex{T}: Multi-task learning with code-text transformer}.
\newblock In \emph{Proceedings of the 1st Workshop on Natural Language Processing for Programming (NLP4Prog 2021)}, pages 40--47, Online. Association for Computational Linguistics.

\bibitem[{Raffel et~al.(2020)Raffel, Shazeer, Roberts, Lee, Narang, Matena, Zhou, Li, and Liu}]{raffel2020exploring}
Colin Raffel, Noam Shazeer, Adam Roberts, Katherine Lee, Sharan Narang, Michael Matena, Yanqi Zhou, Wei Li, and Peter~J. Liu. 2020.
\newblock Exploring the limits of transfer learning with a unified text-to-text transformer.
\newblock \emph{J. Mach. Learn. Res.}, 21(1).

\bibitem[{Rodeghero et~al.(2014)Rodeghero, McMillan, McBurney, Bosch, and D'Mello}]{ir2}
Paige Rodeghero, Collin McMillan, Paul~W. McBurney, Nigel Bosch, and Sidney D'Mello. 2014.
\newblock \href {https://doi.org/10.1145/2568225.2568247} {Improving automated source code summarization via an eye-tracking study of programmers}.
\newblock In \emph{Proceedings of the 36th International Conference on Software Engineering}, ICSE 2014, page 390–401, New York, NY, USA. Association for Computing Machinery.

\bibitem[{Sanh et~al.(2020)Sanh, Debut, Chaumond, and Wolf}]{sanh2020distilbert}
Victor Sanh, Lysandre Debut, Julien Chaumond, and Thomas Wolf. 2020.
\newblock \href {http://arxiv.org/abs/1910.01108} {Distilbert, a distilled version of bert: smaller, faster, cheaper and lighter}.

\bibitem[{Sennrich et~al.(2016)Sennrich, Haddow, and Birch}]{sennrich-etal-2016-neural}
Rico Sennrich, Barry Haddow, and Alexandra Birch. 2016.
\newblock \href {https://doi.org/10.18653/v1/P16-1162} {Neural machine translation of rare words with subword units}.
\newblock In \emph{Proceedings of the 54th Annual Meeting of the Association for Computational Linguistics (Volume 1: Long Papers)}, pages 1715--1725, Berlin, Germany. Association for Computational Linguistics.

\bibitem[{Shi et~al.(2022)Shi, Wang, Du, Chen, Han, Zhang, Zhang, and Sun}]{eval_code_summ}
Ensheng Shi, Yanlin Wang, Lun Du, Junjie Chen, Shi Han, Hongyu Zhang, Dongmei Zhang, and Hongbin Sun. 2022.
\newblock \href {https://doi.org/10.1145/3510003.3510060} {On the evaluation of neural code summarization}.
\newblock In \emph{Proceedings of the 44th International Conference on Software Engineering}, ICSE '22, page 1597–1608, New York, NY, USA. Association for Computing Machinery.

\bibitem[{Shinn et~al.(2023)Shinn, Cassano, Labash, Gopinath, Narasimhan, and Yao}]{shinn2023reflexion}
Noah Shinn, Federico Cassano, Beck Labash, Ashwin Gopinath, Karthik Narasimhan, and Shunyu Yao. 2023.
\newblock \href {http://arxiv.org/abs/2303.11366} {Reflexion: Language agents with verbal reinforcement learning}.

\bibitem[{Sieper et~al.(2020)Sieper, Amarkhel, Diez, and Petrak}]{Sieper2020}
Anna~Abad Sieper, Omar Amarkhel, Savina Diez, and Dominic Petrak. 2020.
\newblock Semantic code search with neural bag-of-words and graph convolutional networks.
\newblock In \emph{SKILL 2020 - Studierendenkonferenz Informatik}, pages 103--115, Bonn. Gesellschaft für Informatik e.V.

\bibitem[{Sontakke et~al.(2022)Sontakke, Patwardhan, Vig, Medicherla, Naik, and Shroff}]{sontakke2022code}
Ankita~Nandkishor Sontakke, Manasi Patwardhan, Lovekesh Vig, Raveendra~Kumar Medicherla, Ravindra Naik, and Gautam Shroff. 2022.
\newblock Code summarization: Do transformers really understand code?
\newblock In \emph{Deep Learning for Code Workshop}.

\bibitem[{Sridhara et~al.(2010)Sridhara, Hill, Muppaneni, Pollock, and Vijay-Shanker}]{template1}
Giriprasad Sridhara, Emily Hill, Divya Muppaneni, Lori Pollock, and K.~Vijay-Shanker. 2010.
\newblock \href {https://doi.org/10.1145/1858996.1859006} {Towards automatically generating summary comments for java methods}.
\newblock In \emph{Proceedings of the 25th IEEE/ACM International Conference on Automated Software Engineering}, ASE '10, page 43–52, New York, NY, USA. Association for Computing Machinery.

\bibitem[{Touvron et~al.(2023)Touvron, Martin, Stone, Albert, Almahairi, Babaei, Bashlykov, Batra, Bhargava, Bhosale, Bikel, Blecher, Ferrer, Chen, Cucurull, Esiobu, Fernandes, Fu, Fu, Fuller, Gao, Goswami, Goyal, Hartshorn, Hosseini, Hou, Inan, Kardas, Kerkez, Khabsa, Kloumann, Korenev, Koura, Lachaux, Lavril, Lee, Liskovich, Lu, Mao, Martinet, Mihaylov, Mishra, Molybog, Nie, Poulton, Reizenstein, Rungta, Saladi, Schelten, Silva, Smith, Subramanian, Tan, Tang, Taylor, Williams, Kuan, Xu, Yan, Zarov, Zhang, Fan, Kambadur, Narang, Rodriguez, Stojnic, Edunov, and Scialom}]{touvron2023llama}
Hugo Touvron, Louis Martin, Kevin Stone, Peter Albert, Amjad Almahairi, Yasmine Babaei, Nikolay Bashlykov, Soumya Batra, Prajjwal Bhargava, Shruti Bhosale, Dan Bikel, Lukas Blecher, Cristian~Canton Ferrer, Moya Chen, Guillem Cucurull, David Esiobu, Jude Fernandes, Jeremy Fu, Wenyin Fu, Brian Fuller, Cynthia Gao, Vedanuj Goswami, Naman Goyal, Anthony Hartshorn, Saghar Hosseini, Rui Hou, Hakan Inan, Marcin Kardas, Viktor Kerkez, Madian Khabsa, Isabel Kloumann, Artem Korenev, Punit~Singh Koura, Marie-Anne Lachaux, Thibaut Lavril, Jenya Lee, Diana Liskovich, Yinghai Lu, Yuning Mao, Xavier Martinet, Todor Mihaylov, Pushkar Mishra, Igor Molybog, Yixin Nie, Andrew Poulton, Jeremy Reizenstein, Rashi Rungta, Kalyan Saladi, Alan Schelten, Ruan Silva, Eric~Michael Smith, Ranjan Subramanian, Xiaoqing~Ellen Tan, Binh Tang, Ross Taylor, Adina Williams, Jian~Xiang Kuan, Puxin Xu, Zheng Yan, Iliyan Zarov, Yuchen Zhang, Angela Fan, Melanie Kambadur, Sharan Narang, Aurelien Rodriguez, Robert Stojnic, Sergey Edunov, and Thomas
  Scialom. 2023.
\newblock \href {http://arxiv.org/abs/2307.09288} {Llama 2: Open foundation and fine-tuned chat models}.

\bibitem[{Vaswani et~al.(2017)Vaswani, Shazeer, Parmar, Uszkoreit, Jones, Gomez, Kaiser, and Polosukhin}]{NIPS2017_3f5ee243}
Ashish Vaswani, Noam Shazeer, Niki Parmar, Jakob Uszkoreit, Llion Jones, Aidan~N Gomez, \L~ukasz Kaiser, and Illia Polosukhin. 2017.
\newblock \href {https://proceedings.neurips.cc/paper_files/paper/2017/file/3f5ee243547dee91fbd053c1c4a845aa-Paper.pdf} {Attention is all you need}.
\newblock In \emph{Advances in Neural Information Processing Systems}, volume~30. Curran Associates, Inc.

\bibitem[{Wan et~al.(2019)Wan, Shu, Sui, Xu, Zhao, Wu, and Yu}]{wan2019}
Yao Wan, Jingdong Shu, Yulei Sui, Guandong Xu, Zhou Zhao, Jian Wu, and Philip~S. Yu. 2019.
\newblock \href {https://doi.org/10.1109/ASE.2019.00012} {Multi-modal attention network learning for semantic source code retrieval}.
\newblock In \emph{Proceedings of the 34th IEEE/ACM International Conference on Automated Software Engineering}, ASE '19, page 13–25. IEEE Press.

\bibitem[{Wang et~al.(2021{\natexlab{a}})Wang, Wang, Mi, Zhou, Wan, Liu, Li, Wu, Liu, and Jiang}]{wang2021syncobert}
Xin Wang, Yasheng Wang, Fei Mi, Pingyi Zhou, Yao Wan, Xiao Liu, Li~Li, Hao Wu, Jin Liu, and Xin Jiang. 2021{\natexlab{a}}.
\newblock Syncobert: Syntax-guided multi-modal contrastive pre-training for code representation.
\newblock \emph{arXiv preprint arXiv:2108.04556}.

\bibitem[{Wang et~al.(2022{\natexlab{a}})Wang, Dong, Lu, and Zhou}]{wang2022gypsum}
Yu~Wang, Yu~Dong, Xuesong Lu, and Aoying Zhou. 2022{\natexlab{a}}.
\newblock \href {https://doi.org/10.1145/3524610.3527903} {Gypsum: learning hybrid representations for code summarization}.
\newblock In \emph{Proceedings of the 30th IEEE/ACM International Conference on Program Comprehension}, ICPC '22, page 12–23, New York, NY, USA. Association for Computing Machinery.

\bibitem[{Wang et~al.(2021{\natexlab{b}})Wang, Wang, Joty, and Hoi}]{wang-etal-2021-codet5}
Yue Wang, Weishi Wang, Shafiq Joty, and Steven~C.H. Hoi. 2021{\natexlab{b}}.
\newblock \href {https://aclanthology.org/2021.emnlp-main.685} {{C}ode{T}5: Identifier-aware unified pre-trained encoder-decoder models for code understanding and generation}.
\newblock In \emph{Proceedings of the 2021 Conference on Empirical Methods in Natural Language Processing}, pages 8696--8708, Online and Punta Cana, Dominican Republic. Association for Computational Linguistics.

\bibitem[{Wang et~al.(2022{\natexlab{b}})Wang, Cuenca, Zhou, Xu, and Neubig}]{mconala}
Zhiruo Wang, Grace Cuenca, Shuyan Zhou, Frank~F. Xu, and Graham Neubig. 2022{\natexlab{b}}.
\newblock \href {https://doi.org/10.48550/ARXIV.2203.08388} {Mconala: A benchmark for code generation from multiple natural languages}.

\bibitem[{Wang et~al.(2023)Wang, Cuenca, Zhou, Xu, and Neubig}]{wang-etal-2023-mconala}
Zhiruo Wang, Grace Cuenca, Shuyan Zhou, Frank~F. Xu, and Graham Neubig. 2023.
\newblock \href {https://doi.org/10.18653/v1/2023.findings-eacl.20} {{MC}o{N}a{L}a: A benchmark for code generation from multiple natural languages}.
\newblock In \emph{Findings of the Association for Computational Linguistics: EACL 2023}, pages 265--273, Dubrovnik, Croatia. Association for Computational Linguistics.

\bibitem[{Wu et~al.(2021)Wu, Zhao, and Zhang}]{wu-etal-2021-code}
Hongqiu Wu, Hai Zhao, and Min Zhang. 2021.
\newblock \href {https://doi.org/10.18653/v1/2021.findings-acl.93} {Code summarization with structure-induced transformer}.
\newblock In \emph{Findings of the Association for Computational Linguistics: ACL-IJCNLP 2021}, pages 1078--1090, Online. Association for Computational Linguistics.

\bibitem[{Yang et~al.(2019)Yang, Dai, Yang, Carbonell, Salakhutdinov, and Le}]{NEURIPS2019_dc6a7e65}
Zhilin Yang, Zihang Dai, Yiming Yang, Jaime Carbonell, Russ~R Salakhutdinov, and Quoc~V Le. 2019.
\newblock \href {https://proceedings.neurips.cc/paper/2019/file/dc6a7e655d7e5840e66733e9ee67cc69-Paper.pdf} {{XLNet}: Generalized autoregressive pretraining for language understanding}.
\newblock In \emph{Advances in Neural Information Processing Systems}, volume~32. Curran Associates, Inc.

\bibitem[{Zhang et~al.(2019{\natexlab{a}})Zhang, Wang, Zhang, Sun, Wang, and Liu}]{zhang2019}
Jian Zhang, Xu~Wang, Hongyu Zhang, Hailong Sun, Kaixuan Wang, and Xudong Liu. 2019{\natexlab{a}}.
\newblock \href {https://doi.org/10.1109/ICSE.2019.00086} {A novel neural source code representation based on abstract syntax tree}.
\newblock In \emph{2019 IEEE/ACM 41st International Conference on Software Engineering (ICSE)}, pages 783--794.

\bibitem[{Zhang et~al.(2019{\natexlab{b}})Zhang, Kishore, Wu, Weinberger, and Artzi}]{bertscore}
Tianyi Zhang, Varsha Kishore, Felix Wu, Kilian~Q. Weinberger, and Yoav Artzi. 2019{\natexlab{b}}.
\newblock \href {http://arxiv.org/abs/1904.09675} {Bertscore: Evaluating text generation with {BERT}}.
\newblock \emph{CoRR}, abs/1904.09675.

\end{thebibliography}

\appendix
\label{sec:appendix}

\section{Appendix}

\subsection{Distribution of BLEU Scores}
\label{appendix:dist_bleu}

\begin{figure}[h!]
\centering
  \includegraphics[width=0.5\textwidth]{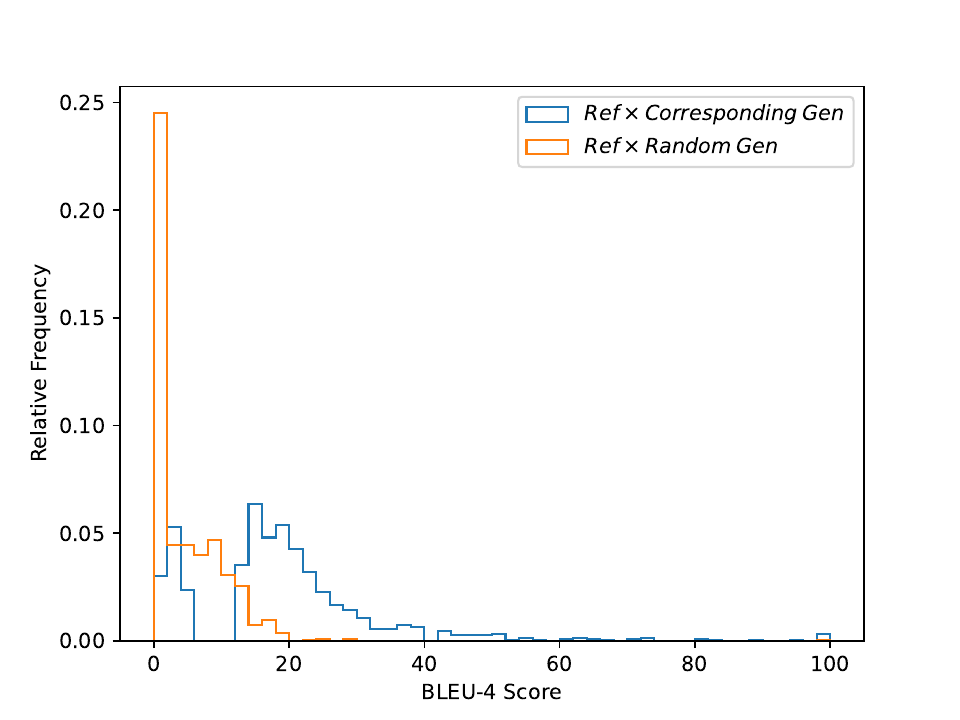}
  \caption{Distribution of BLEU scores between reference descriptions and their corresponding generated descriptions, and between reference descriptions and a generated description from another random example across all models and variants}
  \label{fig:bleu_corresponding_vs_random}
\end{figure}

Similar to BERTScores in Section \ref{sec:bertscore_dist}, we plot the distribution of BLEU scores between random pairs of reference descriptions and generated descriptions, as well as between the reference descriptions and their corresponding descriptions in Figure \ref{fig:bleu_corresponding_vs_random}. We see that unlike BERTscores, generated summaries that are from a different example are assigned very low, mostly zero scores. This shows that BLEU can not only discern between a correct and incorrect description better than BERTScore but can also identify an incorrect generation by assigning it a zero score.

\begin{figure}[h!]
\centering
  \includegraphics[width=0.5\textwidth]{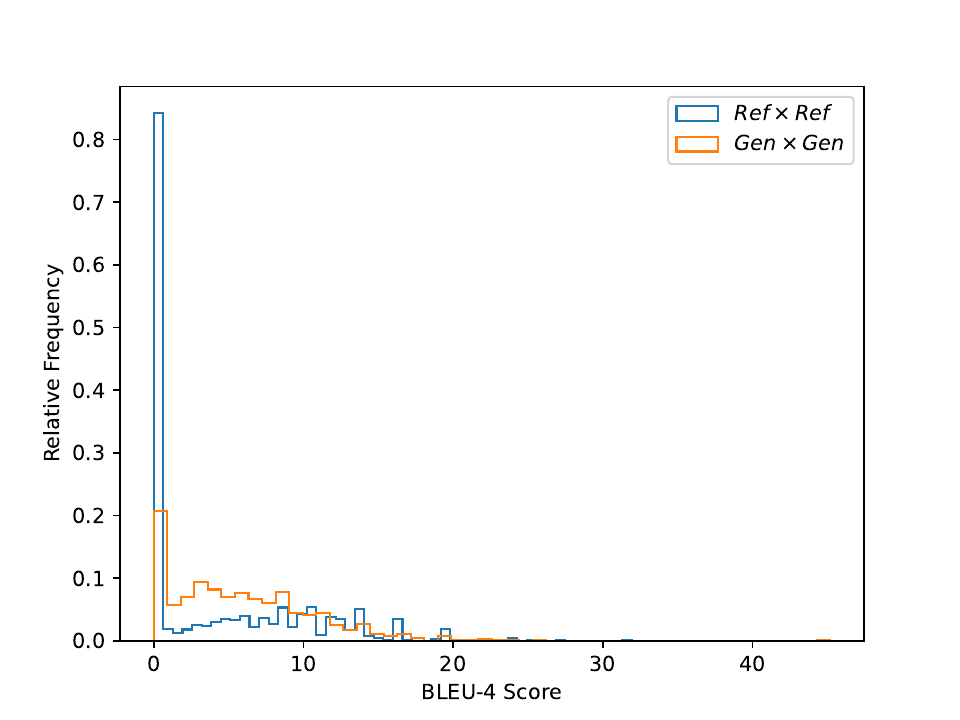}
  \caption{Distribution of BLEU Scores between two random reference descriptions and two random generated descriptions across all models and variants}
  \label{fig:bleu_reference_vs_generated}
\end{figure}

In Figure \ref{fig:bleu_reference_vs_generated}, we see that the distribution of BLEU scores between two randomly sampled generated descriptions is higher than those between two randomly sampled reference descriptions, showing that the models have a problem of producing generations that are less diverse than the data they were trained on. We made a similar observation in Section \ref{sec:bertscore_dist} which shows that the problem persists independent of the metric used.

\end{document}